\documentclass[11pt,letter]{article}
\usepackage{graphicx}
\usepackage{amssymb}
\input epsf

\newcommand{\be}{\begin{equation}}
\newcommand{\beq}{\begin{equation}}
\newcommand{\bea}{\begin{eqnarray}}
\newcommand{\ee}{\end{equation}}
\newcommand{\eeq}{\end{equation}}
\newcommand{\eea}{\end{eqnarray}}
\newcommand{\tr}{{\rm Tr}\,}
\newcommand{\re}{{\rm Re}\,}
\newcommand{\lp}{(}
\newcommand{\rp}{)}
\newcommand{\ie}{i.e.}
\newcommand{\tom}{\tilde{\omega}}
\newcommand{\toom}{\tilde{\Omega}}
\newcommand{\tx}{\tilde{x}}

\newcommand{\pd}{\partial}

\title{\boldmath  On the Markov evolution of the $\rho$-matrix of a subsystem.}

\author{M.A.Braun\\
{\it Dep. High-energy physics, Saint-Petersburg State University, Russia}}

\begin{document}
\maketitle
\begin{abstract}
Evolution of the reduced density matrix for a subsystem is studied to determine deviations from its Markov character
for a system consisting of  a closed chain of $N$ oscillators with one of them serving as a subsystem. The dependence on $N$ and on the coupling of the
two subsystems is investigated numerically. The found deviations strongly depend on $N$ and the coupling. In the most beneficial case with $N-1=100$
and the coupling  randomized in its structure the deviations fall with the evolution time up 3\%. In other cases they remain to be of the order
30\% or even more.
\end{abstract}

\section{Introduction}
If the whole system is split in two interacting subsystems, 0 and 1, the subsystem 0 under investigation (observable) and the other 1 as "the bath",
one defines the reduced density matrix $\rho_0$ as a partial trace of the total one $\rho$ over the bath variables. The time evolution of the reduced
density matrix $\rho_0$ is of course uniquely determined by the evolution of the total density matrix and to some degree should depend on the evolution of the "bath".
It is customary to describe the time evolution of the reduced density matrix by the Franke-Lindblad-Gorini-Kossakowsky-Sudarshan (FLGKS) equation
 ~\cite{franke,lindblad,lindblad1,gorini}. This equation is a consequence of the
four properties assumed for the evolution: conservation of unit trace, non-negativity of diagonal matrix elements, linearity and markovian character.
It is the latter property which finally allows to derive the FLGKS equation (see ~\cite{pearle,brasil,lidar2001,lidar,lidar1}). And of course this is the least convincing assumption. In simple words it
allows to predict the reduced matrix $\rho_0(t)$ at time $t$ provided it is known at some previous time $t_0$. Mathematically it implies that the derivative in time
$d\rho_0(t)/dt$ is expressed by the value of $\rho_0(t)$ itself and not by some integral of $\rho_0(t')$ over intermediate times $t'$, $t_0<t'<t$.
In fact this assumption looks rather doubtful from the start. Many authors have tried to derive  this markovianity from certain specific properties of the bath
(see e.g \cite{brasil} and lecture notes ~\cite{lidar}). All these derivations actually include  additional assumptions together with certain particular properties of the
bath. In fact in these papers the attention was centered on deriving the FLKS equation or its certain generalizations for the condensed matter applications with the baths
corresponding to a more or less realistic macroscopic and time dependent surroundings.

In this note we abandon this purposeful approach and attempt a mini-investigation of the markovianity itself for a bath of finite or even small
dimensions. We study a more abstract problem of splitting a system in two without bothering about their degrees of freedom. As a simplest
case we study  the exactly soluble model of $N$   interacting one-dimensional
oscillators one of which is our observed system 0 and the other $N-1\equiv N_1$ from 1 to $ N_1$ serve as the bath (system 1). In this case both $\rho(t)$ and $\rho_0(t)$ are explicitly known.
For the Markov evolution, once we know the reduced density matrix $\rho_0(t_0)$ at time $t_0$,  its value at time $t$ should be uniquely determined, that is $\rho_0(t)$
is fully determined by $\rho_0(t_0)$.  Any change in the initial total $\rho$ matrix at time $t_0$ which varies the state of system 1 but
preserves the initial value of the reduced matrix $\rho_0(t_0)$ should not change $\rho_0(t)$ at time $t>t_0$.

Our study  shows that of course generally this is  not the case. The reduced density matrix at time $t$ is not uniquely determined by its value at $t_0<t$ but
depends on the initial state of the bath, clearly manifesting non-markovianity of its evolution.
This dependence depends on the time elapsed in the course of evolution, on the properties of the bath, in particular on the number $N-1$ of its oscillators, and
on the coupling between the system and bath. For the latter the crucial role is played by the sort of interaction between the oscillator 0 and the rest.
In the most beneficial case when this coupling resembles what is usually implied under system-bath coupling and when there are no exclusive parts of the bath connected to the
system, we find  deviations from markovianity steadily diminishing with time. In this case,
if $T$ is the average characteristic time of the bath (that is the time at which the bath strongly varies) then with $N=100$ at evolution times $t$ of the order $t/T\sim 50$
the deviation from markovianity falls to 3\%. However for other sort of couplings and of course with a small bath of only 2 oscillators this is not the case
and deviations  of markovianity stay at the order 30\% or even more.

\section{Model: a lattice of oscillators}
We use the model borrowed from~\cite{jefferson}. In this section we
just recapitulate the main points necessary for the following.

Take  $N$ oscillators on a one-dimensional circular lattice,
\be
H=\frac{1}{2}\sum_{a=0}^{N_1=N-1}\left[  p_a^2+\omega^2\, x_a^2+\lambda\lp x_a-x_{a+1}\rp^2\right]~,
\label{qm88}
\ee
with periodic boundary conditions $x_{a+N}=x_a$. Here $\hbar=1$ and $\omega$ has the units of energy.
In the calculations we put $\omega=1$.
To solve the above system, one  rewrites the Hamiltonian in terms of the normal modes,
\be
H=\frac{1}{2}\sum_{k=0}^{N_1}\left[ \, |\tilde p_k|^2+\tom_k^2\  |\tilde x_k|^2\,\right]~,
\label{qm288}
\ee
where the transformation to the normal-mode basis is achieved by a (discrete) Fourier transform,
\beq
\tilde x_k\equiv \frac{1}{\sqrt{N}}\sum_{a=0}^{N_1}\mathrm{exp}\lp-\frac{2\pi i\,k}{N}\,a\rp x_a~.\label{eq:Fourier}
\eeq
The normal mode  momenta are defined as
\beq
\tilde p_k\equiv \frac{1}{\sqrt{N}}\sum_{a=0}^{N_1}\mathrm{exp}\lp\frac{2\pi i\,k}{N}\,\,a\rp p_a~.
\eeq
They satisfy
the standard commutation relations:
$[\tilde x_k,\tilde p_{k'}]=i\delta_{kk'}$ and $[\tilde x_k,\tilde x_{k'}]=0=[\tilde p_k,\tilde p_{k'}]$.
The normal-mode frequencies $\tom_k$ are found in terms of the physical frequencies $\omega$ and $\lambda$ in the Hamiltonian (\ref{qm88}) as follows
~\cite{jefferson}:
\beq
\tom_k^2=\omega^2+4\lambda\,\sin^2\!\frac{\pi k}{N}. \label{eq:eigenfreq}
\eeq
Equation (\ref{qm288}) reduces the whole system  to the set of $N$ decoupled harmonic oscillators, with the ground-state wave function
\beq
\psi_0(\tilde x_0,\tilde x_1, \tilde x_2,\cdots)
=\prod_{k=0}^{N_1}\,\left(\frac{\tom_k }{\pi}\right)^{1/4}\ \mathrm{exp}\!\left[-\frac{1}{2}\,\tom_k\,|\tilde x_k|^2\right]~.\label{targetk}
\eeq

To express the wave function in terms of the original variables $x_a$ in the position basis one
writes the Fourier transformation (\ref{eq:Fourier}) between the position and normal-mode bases as $\tilde x = S\, x$, with
\beq
S=\frac{1}{\sqrt{N}}
\left(\begin{array}{ccccc}
1 &  1 & 1 & \ldots & 1 \\
1 & \mu & \mu^2 & \ldots & \mu^{N-1} \\
1 & \mu^2 & \mu^4 & \ldots & \mu^{2(N-1)} \\
\vdots & \vdots & \vdots & \ddots & \vdots \\
1 & \mu^{N-1} & \mu^{2(N-1)} & \ldots & \mu^{(N-1)^2}
\end{array}\right)
\label{logan}
\eeq
where $\mu\equiv\exp\lp-2\pi i/N\rp$.
Since $S$ is a unitary matrix, \ie  $S^\dagger S=1$,
the inverse transformation is given by $x =S^\dagger\,\tilde x$.
In particular the ground state (\ref{targetk}) is rewritten as
\beq
\psi_0(x_a)=\prod_{k=0}^{N_1}\lp\frac{\omega_k}{\pi}\rp^{\frac14}\ \exp\left[-\frac12\, x^T A_T\,  x \right]\ \
{\rm with}\ \  A_T=S^\dagger \tilde A_T\,S\ \  {\rm and}\ \  \tilde A_T=\mathrm{diag}\lp\tom_0,\ldots,\tom_{N-1}\rp~.
\label{targetX}
\eeq
The relation $\tom_k=\tom_{N-k}$ ensures that $A_T$ is real.

Note that mode variables $\tx$ are complex. Setting $\tx=\xi+i\eta$ and $\tx^*=\xi-i\eta$ one finds that in terms of real variables
our system consists of two identical copies of $N$ oscillators with mode coordinates $\xi$ and $\eta$.

\section{Time evolution}
In terms of real variables we find
\be
H=\frac{1}{2}\sum_{k=0}^{N_1}\Big(\frac{\pd^2}{\pd\xi_k^2}+\frac{\pd^2}{\pd\eta_k^2}+\tom_k^2(\xi^2+\eta^2)\Big),
\ee
from which we conclude that the oscillators in $\xi$ and $\eta$ evolve independently. In particulary the evolution of the oscillators $\xi$
is realized by the product of oscillator Green functions
\be
\prod_{k=0}^{N_1}\Big(\frac{g_k}{2\pi i}\Big)^{1/2}\exp \Big\{\frac{i}{2}[f_k(\xi_k^2+{\xi_k'}^2)-2g_k\xi_k\xi_k']\Big\}
\ee
with $f_k=\tom_k\cot(\tom_kt)$ and $g_j=\tom_k/\sin(\tom_kt)$.
The evolution of the oscillators $\eta$ will be achieved by the same product with $\xi,\xi'\to\eta\eta'$.

We take our initial wave function as a Gaussian:
\be
\psi(x)=\Big(\frac{\det(\re\,\Omega)}{\pi^2}\Big)^{1/4}\exp\Big(-\frac{1}{2}(x\Omega x)\Big),
\label{psi0}
\ee
where $(x\Omega x)=\sum_{jk}x_j\Omega_{jk}x_k$  and $\Omega$ is some matrix which characterizes the initial state
of the whole system. Using $\tx=Sx$ and so $x=S^{-1}\tx$ where $S$ is unitary we transform (\ref{psi0}) to
variables $\tx$
\be
\psi(\tx)=\Big(\frac{\det(\re\,\toom)}{\pi^2}\Big)^{1/4}\exp\Big(-\frac{1}{2}(\tx^\dagger\toom \tx)\Big),
\label{psi00}
\ee
where $\toom=S\Omega S^{-1}$. In terms of real variables
\[(\tx^\dagger\toom\tx)=(\xi-i\eta|\toom|\xi+i\eta)=(\xi\toom\xi)+(\eta\toom\eta)-i(\eta\toom\xi)+i(\xi\toom\eta).\]
Due to symmetry of the matrix $\toom$ the two last terms cancel and we find
\be
\psi(\tx)=\Big(\frac{\det(\re\,\toom)}{\pi^2}\Big)^{1/4}\exp\Big(-\frac{1}{2}(\xi\toom \xi)-\frac{1}{2}(\eta\toom\eta)\Big),
\label{psi01}
\ee
which is the product of two independent wave functions for oscillators $\xi$ and $\eta$. So they will involve independently.

Consider the evolution of oscillators $\xi$. We have
\[\psi(\xi,t)=\Big(\frac{\det(\re\,\toom)}{\pi^2}\Big)^{1/4}
\prod_{k=0}^{N_1}\Big(\frac{g_k}{2\pi i}\Big)^{1/2}\exp\Big(\frac{i}{2}f_k\xi_k^2\Big)\]\[\times
\int d \xi'_k\exp\Big(\frac{i}{2}[f_k{\xi'}^2-ig_k\xi_k\xi'_k]-\frac{1}{2}(\xi'\toom\xi')\Big).\]
The exponent $P$ of the last exponential can be presented as
\be
P=\sum_{j,k}\Big(-\frac{1}{2}\xi'_j(\toom_{jk}-if_k\delta_{jk})\xi'_k \Big)\equiv \Big(\xi'(\toom -if)\xi'\Big),
\eeq
where $f_{jk}\equiv f_k\delta_{jk}$. The Gaussian $N$ dimensional integral over $\xi'_k$ gives
\be
\int\prod_k d\xi'_k e^P=\Big(\frac{\pi^N}{\det(\toom-if)}\Big)^{1/2}\exp\Big[\frac{1}{2}\Big( g(\toom -if)^{-1}g\Big)\Big],
\eeq
where $g_{jk}=g_k\delta_{jk}$. Taking into account the exponential outside the integral we find in the end
\be
\psi(\xi)=C\exp\Big[-\frac{1}{2}\Big(\xi\toom(t)\xi\Big)\Big]
\label{psit}
\ee
where
\be
\toom(t)=g(\toom(0)-if)^{-1}g-if
\eeq
and we designated the initial $\toom$ in (\ref{psi01}) as $\toom(0)$.

The evolution of $\psi(\eta,t)$ will give a similar expression with $\xi\to\eta$.
Combining these two results we shall find in variables $\tx$
\be
\psi(\tx,t)=\Big(\frac{\det(\re\,\toom(t))}{\pi^2}\Big)^{1/4}\exp\Big[-\frac{1}{2}\Big(\tx^\dagger\toom(t)\tx\Big)\Big].
\ee
The normalization coefficient is obvious from the fact that the wave function remains correctly normalized during the time evolution.

Now we return to our initial real variables $x$. To do this it is necessary to rotate our matrices with the transformation $S$.
So we finally find
\be
\psi(x,t)=\Big(\frac{\det(\re\,\Omega(t))}{\pi^2}\Big)^{1/4}\exp\Big[-\frac{1}{2}\Big(x\Omega(t)x\Big))\Big],
\ee
where
\be
\Omega(t)=G(\Omega(0)-iF)G-iF
\label{omegat}
\ee
with
\be
F=S^{-1}fS,\ \ G=S^{-1}gS
\ee
and we designated the initial matrix $\Omega$ in (\ref{psi0}) as $\Omega(0)$.

\section{ Density Matrix}

In the position representation
the $\rho$-matrix as an operator is constructed from the wave function and its complex conjugate.
\be
\hat{\rho}=\int\prod_{i=0}^{N_1}dx_idx'_i\rho(x, x')|x><x'|
\ee
where $x$ and $x'$ are $N$-dimensional vectors $x=\{x_0,x_1,...x_N\}$ and similarly for $x'$.
At  at the initial time
\be
\rho\equiv \rho( x,x')=\psi(x))\psi^*(x')=
\left ( \frac{ {\rm det}\,{\rm Re}\, \Omega ) }{ \pi^2 } \right)^{1/2}
  \exp \left ( - \frac{1}{2} (x \Omega x)- \frac{1}{2}(x' \Omega x')^* \right ) \
\label{rho0}
\ee
and at time $t$ the $\rho$-matrix will be
\be
\rho'=\rho'( x,x')=\psi(x,t)\psi^*(x',t)=
\left ( \frac{{\rm det}\,{\rm Re}\, \Omega' ) }{ \pi^2 } \right)^{1/2}
  \exp \left ( - \frac{1}{2} (x\Omega' x)- \frac{1}{2}( x'\Omega' x')^* \right ) \ ,
\label{rhot}
\ee
where $\Omega'\equiv \Omega(t)$ is determined from $\Omega\equiv\Omega(0)$ by (\ref{omegat}).
Of course we have the standard property
\[\tr\hat{\rho}=\tr {\hat{\rho}}^2=1.\]

We will be interested in analyzing the reduced density matrix.
To that end, we partition the system into two subsystems, taking oscillator 0 to be
our ``system" and oscillators $1\ldots N-1\equiv N_1$ to be the ``bath".
We form the reduced density matrix of oscillator 0 as
\begin{equation}
\rho_0 = {\rm Tr}_{1\ldots N_1}\,\rho  \ ,
\end{equation}
where $\rho$ is the density matrix of the entire system.

In the position representation
\begin{equation}
 \hat{\rho}_0 = \int dx_1 dx_1'~ \rho_0(x_0,x_0') ~ | x_0 \rangle \langle x'_0 |  \ ,
\end{equation}
where
\begin{equation}
 \rho_0(x_0,x_0') = \int \Big(\prod_{j=1}^{N_1}dx_j\Big)~ \rho( x_0,x_1,..,x_{N_1}\mid x_0',x_1...,x_{N_1})
\label{rhoreduced}
\end{equation}

To do the integrations we separate variables $x_1,...,x_{N_1}$ in the exponent of (\ref{rho0}) with $\bar{x}_k=x_k$ for $k=1,...N_1$.
We introduce a vector $v$ and matrix $M$ of dimensions $N_1$ and $N_1\times N_1$ respectively
\beq
v_k=\Omega_{0k},\ \ M_{jk}=\Omega_{jk},\ \ j,k=1,...,N_1
\eeq
($M$ is the  minor 00  of $\Omega$). We also denote the coordinates of the bath $\bar{x}_k=x_k,\ \  k=1,...,N_1$.
Then
\[ - \frac{1}{2} (x \Omega x)- \frac{1}{2}(x' \Omega x')^*=-\frac{1}{2}\Omega_{00}x_0^2-\frac{1}{2}\Omega^*_{00}{x'_0}^2
-\bar{x}(vx_0+v^*x_0')-\frac{1}{2}\Big(\bar{x}(M+M^*)\bar{x}\Big).\]
The integrations in (\ref{rhoreduced}) then give
 at the initial time
\be
\rho_1(x_0,x'_0)=\sqrt{\frac{{\rm Re}\,R_{11}-R_{12}}{\pi}}\exp\Big\{-\frac{1}{2}x_0^2R_{11}-\frac{1}{2}{x_0'}^2 R_{11}^*+x_0x'_0R_{12}\Big\},
\label{rho10}
\ee
where
\be R_{11}=\Omega_{00}-a,\ \ R_{12}=|a|,\ \ a=\Big(v(M+M^*)^{-1}v\Big).
\label{r0}
\ee
The coefficient in(\ref{rho10}) can be determined from the obvious property
$\tr\hat{\rho}_0=1$.

The purity $\mu=\tr{\hat{\rho}_0}^2$ is generally less than unity.
We find
\[
\rho_0^2(x_1,x'_1)=\int dx \rho_1(x_1,x)\rho_1(x,x')=
\frac{{\rm Re}\,R_{11}-R_{12}}{\pi}\sqrt{\frac{\pi}{\re R_{11}}}\]
\be
\exp\Big\{-\frac{1}{2}x^2\Big(R_{11}-\frac{R_{12}^2}{2\re R_{11}}\Big)
- \frac{1}{2}{x'}^2\Big(R_{11}^*-\frac{R_{12}^2}{2\re R_{11}}\Big)+xx'\frac{R_{12}^2}{2\re R_{11}}\Big\}.
\ee
Taking the trace we find
\be
\mu=\tr{\hat{\rho}_0}^2=\sqrt{\frac{{\rm Re}\,R_{11}-R_{12}}{{\rm Re}\,R_{11}+R_{12}}}.
\ee

At time $t$ we have the same formulas for the reduced matrix $\rho_0'$ in which $\Omega\to\Omega'$
and $R\to R'$:
\be
\rho'_1(x_0,x'_0)=\sqrt{\frac{\re R'_{11}-R'_{12}}{\pi}}
\exp\Big\{-\frac{1}{2}x_0^2R'_{11}-\frac{1}{2}{x_0'}^2 {R'_{11}}^*+x_0x'_0R'_{12}\Big\},
\label{rho1t}
\ee
where
\be R'_{11}=\Omega'_{00}-a',\ \ R'_{12}=|a'|,\ \ a'=\Big(v'(M'+{M'}^*)^{-1}v'\Big).
\label{rt}
\ee
We recall that $\Omega'$ is determined via $\Omega$ by  (\ref{omegat}). Of course $\tr\hat{\rho}'_0=1$
and the  purity $\mu'$ will be given by
\be
\mu'=\tr[{\hat{\rho}_0}']^2=\sqrt{\frac{\re R'_{11}-R'_{12}}{\re R'_{11}+R'_{12}}}.
\ee

Equations (\ref{rho10}) -- (\ref{rt}) describe evolution of the reduced density matrix in time.
In the next section we shall study its properties and whether it is Markovian or not and if not to what degree at different times and properties of system 1.

\section{Time evolution of the reduced density matrix}
The density matrix $\rho_0$ is fully determined by the 2$\times$ 2 matrix $\hat{R}=\{R_{11},R_{12},R_{21}=R_{12},R_{22}=R_{11}^*\}$. 
So our strategy will be to fix $\hat{R}$ at the initial time $t_0$ and
study if one can uniquely determine the $\rho_0$ at some later $t$, that is to find in a unique manner its matrix $\hat{R}'$.
If this is possible then the process has the Markovian character. However at time $t$ matrix $\hat{R}'$ is actually expressed via  matrix $\Omega'$,
which is determined by $\Omega$ by the evolution equation for the whole system.
We assume the initial symmetric matrix $\Omega$ to be real. Then with fixed $R_{11}$ and $R_{12}$ we are left with $N(N-1)/2$ parameters in 00-minor
$M$ and (N-1) parameters in $v$ constrained by the value of $\hat{R}$, that is $(N+2)(N-1)/2-1$ parameters in all, which characterize the bath at fixed $R_{11}$ and $R_{12}$.
So it may happen that with  fixed $\hat{R}$ we shall not be able to find $\hat{R}'$ in a unique manner, since the latter will depend on extra variables in $M$ and $v$ left
 after fixing $\hat{R}$ and actually depending on the state of the "bath". In this case the evolution will be non-Markovian.
 The degree of the dependence on the mentioned extra variables will measure the scale of non-Markovianity. It will possibly depend on time, dynamic
 properties of the bath and its coupling to system 0, that is on $t$  and $\lambda$, having in mind that the value of $\omega$ just fixes the energy (and time) units.

Our plan is to study this dependence numerically. We shall choose some initial state of the whole system, that is matrix $\Omega$,
determine some of its matrix elements from fixed matrix $\hat{R}$ and vary the rest of them. Then we shall study the purity $\mu'$ at later time $t$
and see if it changes when $\hat{R}$ is fixed but the rest of matrix elements in $\Omega$ change.

We shall also study how this change depends on the value of $\lambda$ choosing $\lambda=0.1$ or $\lambda=1.0$
in units $\omega=1$

As already mentioned we choose the initial matrix $\Omega$ to be real. We fix the matrix element $\Omega_{00}=1$ at the initial moment $t=0$.
The rest $N_1=N-1$  diagonal matrix elements $\Omega_{kk}$ $k=1,N_1$ at the initial moment will be allowed to take different values, thus characterizing  different
states of the bath with bath parameters $BP$. We shall consider five possible sets of these diagonal matrix elements which
preserve the average state of the bath, that is $\tr \Omega$, and have the bath parameters $BP(i)$, $i=1,...5$.\\
 $ BP(1):\ \ \Omega_{kk}=1,\ \ k=1,...N_1$.\\
 $ BP(2):\ \ \Omega_{11}=1.5,\ \ \Omega_{22}=0.5,\ \ \Omega_{kk}=1,\ \ k=3,...N_1$.\\
 $ BP(3):\ \ \Omega_{N_1/2,N_1/2}=1.5,\ \ \Omega_{N_1/2+1,N_1/2+1}=0.5,\ \ \Omega_{kk}=1\ \ {\rm for\ the\ rest}\ k$.\\
 $ BP(4):\ \ \Omega_{kk}=1.5,\ \ k=1,...N_1/2,\ \ \Omega_{kk}=0.5 \ \ k=N_1/2+1,...N_1$.\\
 $ BP(5):\ \ \Omega_{kk}=1.5\ \ {\rm or}\ \ 0.5$ at randomly chosen equal number of $k$ from $k=1,...N_1$.\\
Our aim will be to study the dependence of the evolution with these varying states of the bath.
Note that these states of the bath are not completely equivalent. The subsystem 0 directly interacts with only oscillators 1 and $N_1$.
So one expects that changing matrix elements $\Omega_{11}$ and $\Omega_{N_1N_1}$ will have greater influence of the  bath state than changing
other matrix elements or randomizing the change. In this sense the idea of the "bath" is better suited to $BP(1)$, $BP(3)$ and $BP(5)$ rather than
to $BP(2)$ and $BP(4)$.

As a convenient signature we take the purity $\tr \rho_0^2(t)$ during evolution.
 In the following we study the bath dependence comparing the purity from the sets $BP(1,3,5)$ and $BP(1,..5)$.

We shall consider two cases with the initial values (at $t=0$)   $R_{12}=0$, $\mu=1$ (case A) and $R_{12}=0.325$, $\mu=0.592$ (case B).
In both cases at the initial moment $R_{11}=\Omega_{00}-R_{12}=1-R_{12}$ and non-diagonal elements $\Omega_{0k}$, $k=1,N_1$
are taken equal and adjusted to the fixed value of $R_{12}$. In particular for $R_{12}=0$ they are all equal to zero.

\subsection{Numerical results, $\lambda=0.1$}

\begin{figure}
\begin{center}
\includegraphics[width=8. cm]{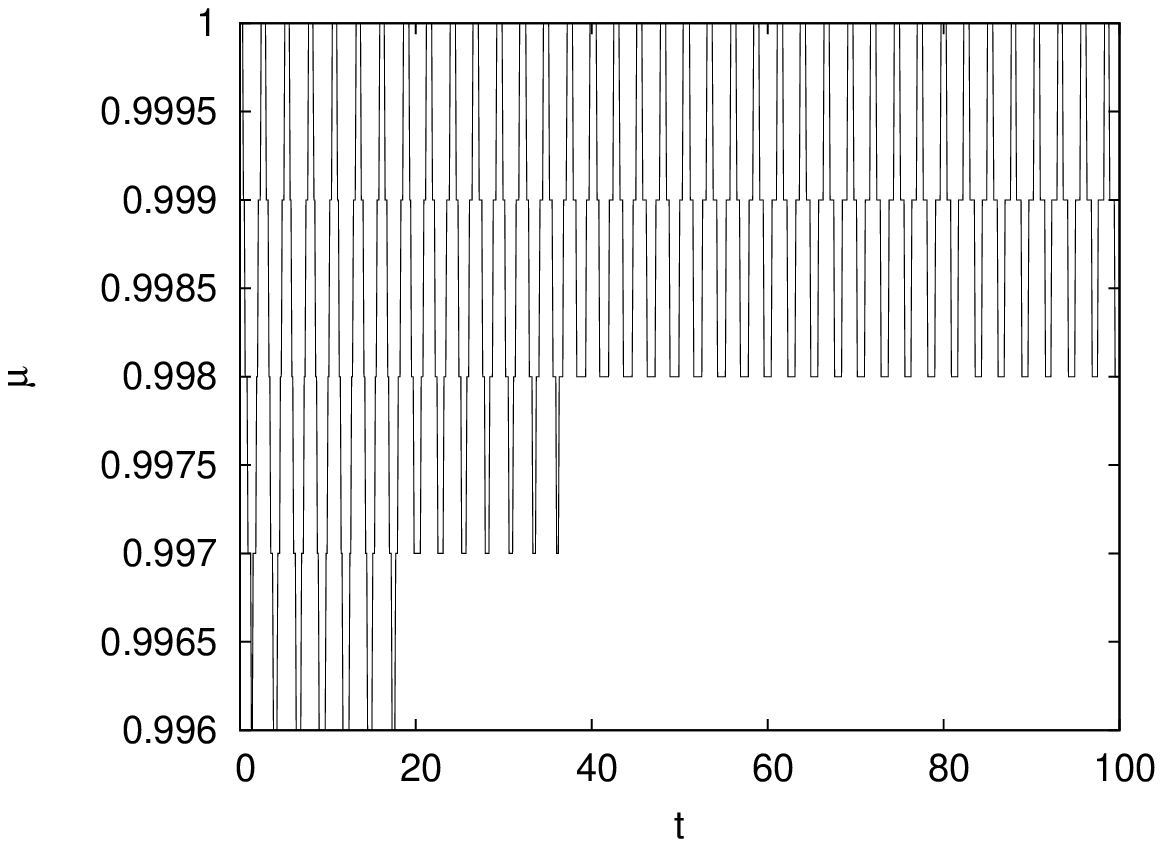}
\includegraphics[width=8. cm]{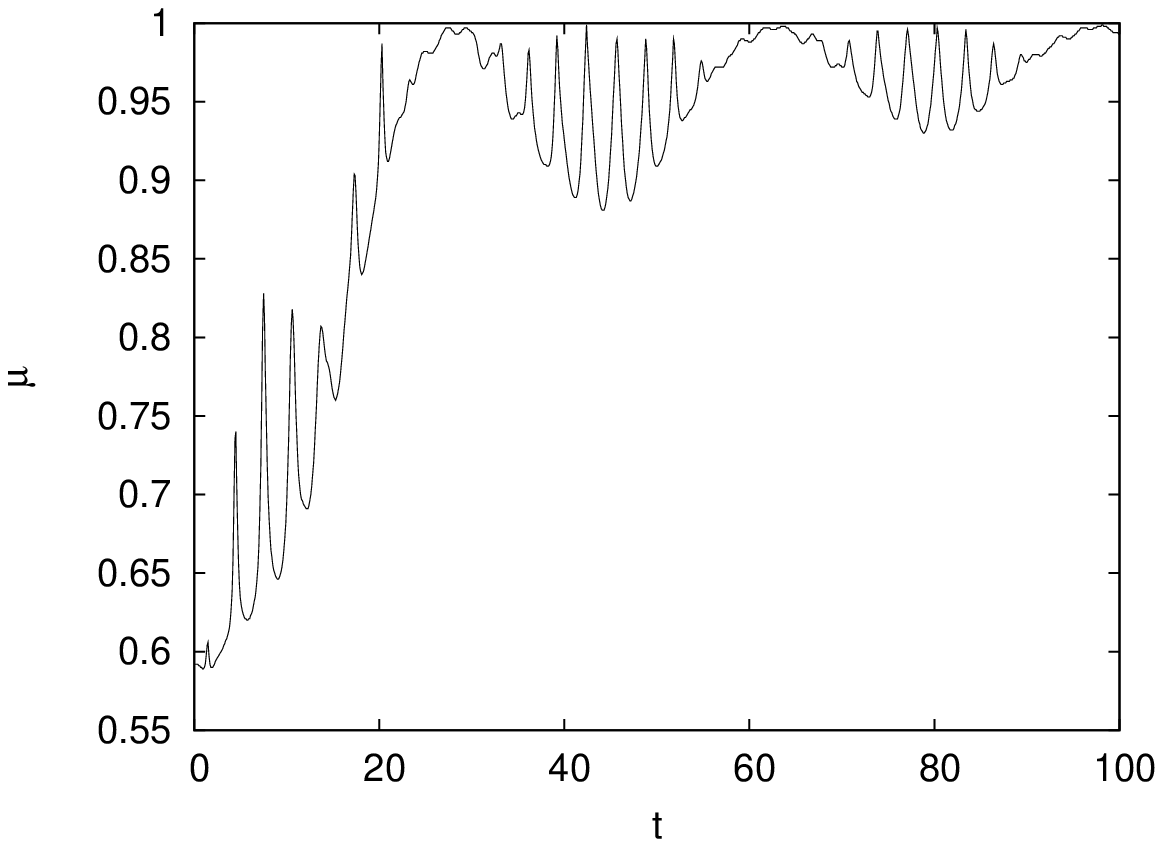}
\caption{The calculated purity $\mu$ at different times $t$ for $\lambda=0.1$, cases A(left panel) and B( right panel),
bath parameter $BP(1)$ and number of oscillators $N=101$}
\label{fig1}
\end{center}
\end{figure}

Note that as a function of $t$ purity $\mu(t)$ in both cases oscillates with frequency $\sim 2.5$. So in the interval from $t=0$ up to $t=100$ the
plots of $\mu(t)$ show  a lot of oscillations. This is illustrated in Fig. \ref{fig1} in which we show natural values of $\mu(t)$ for $BP(1)$ and
both cases A and B with $N_1=100$

Through the maze of oscillation however one can see the general trend of the purity, once  averaged over the short range oscillations.
To make the graphical presentation more convenient in the following we show the purity  averaged over these short range oscillation by
the Bezier interpolation ~\cite{bezier}.\\

\vspace*{0.2 cm}
{\bf Case A}
\vspace*{0.2 cm}

\begin{figure}
\begin{center}
\includegraphics[width=8. cm]{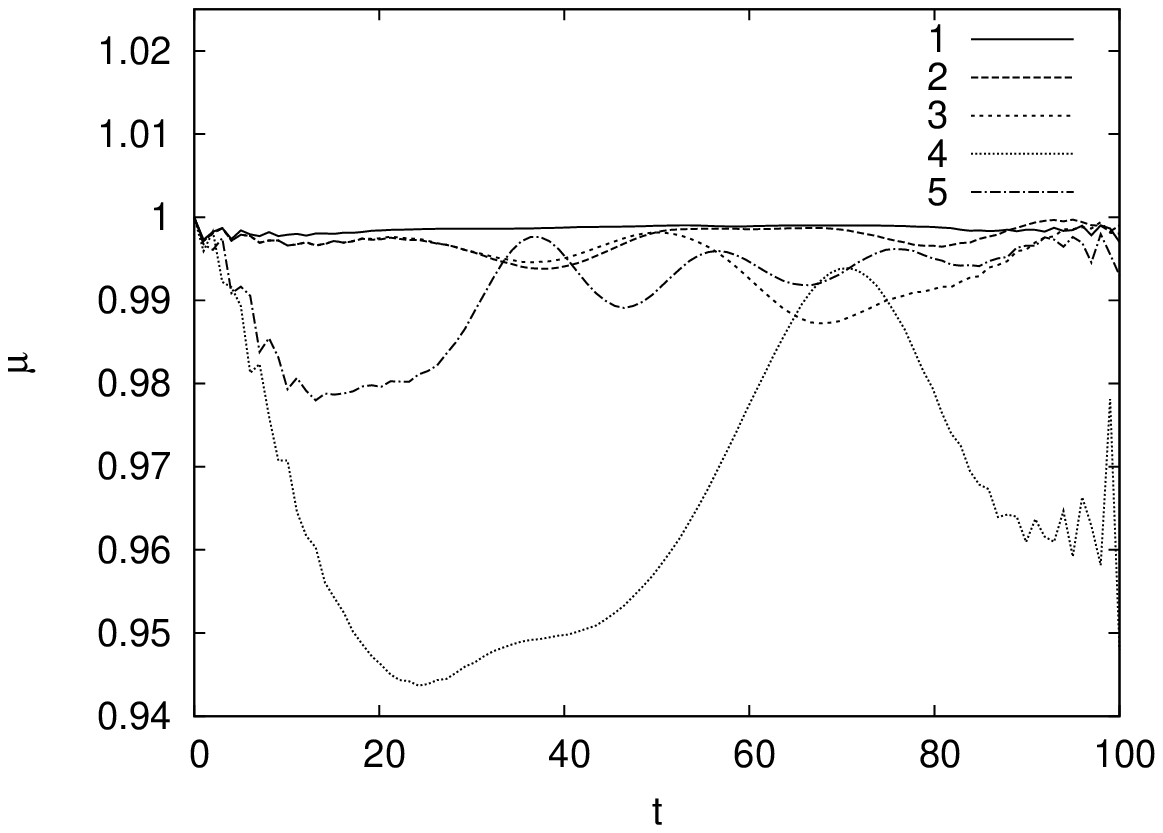}
\includegraphics[width=8. cm]{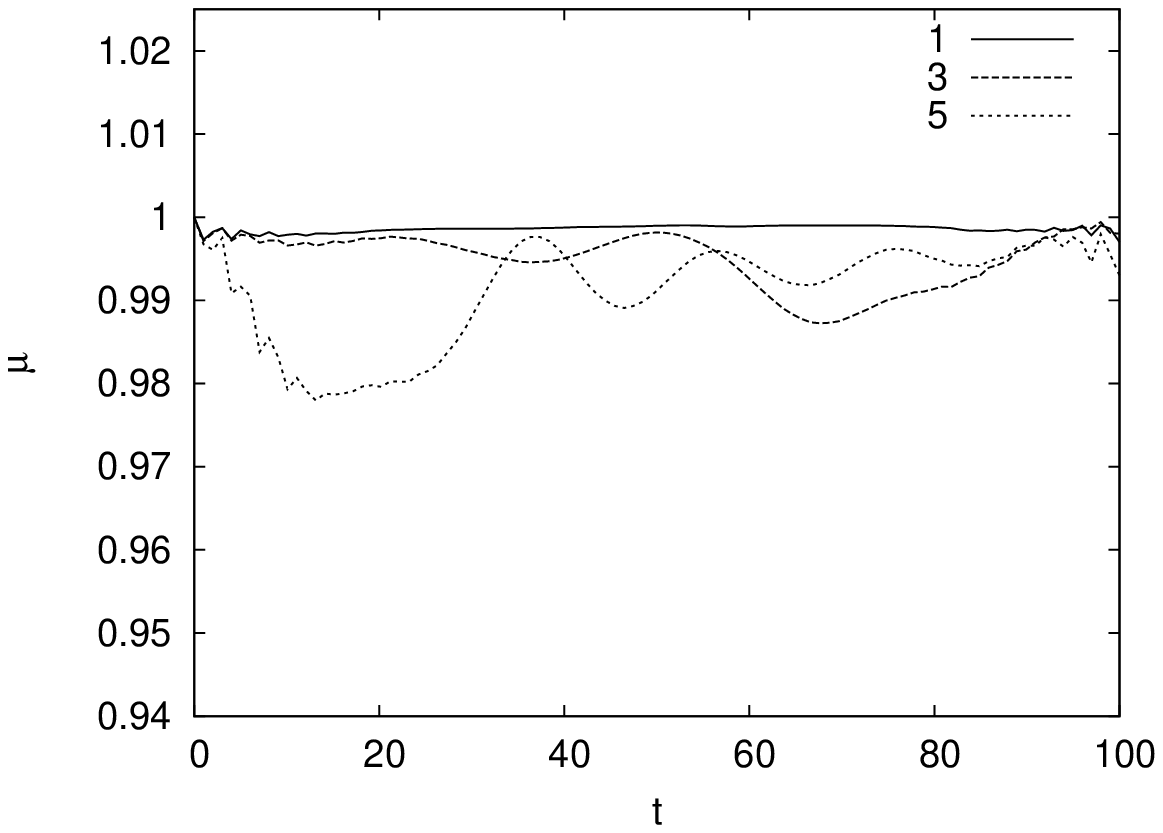}
\includegraphics[width=8. cm]{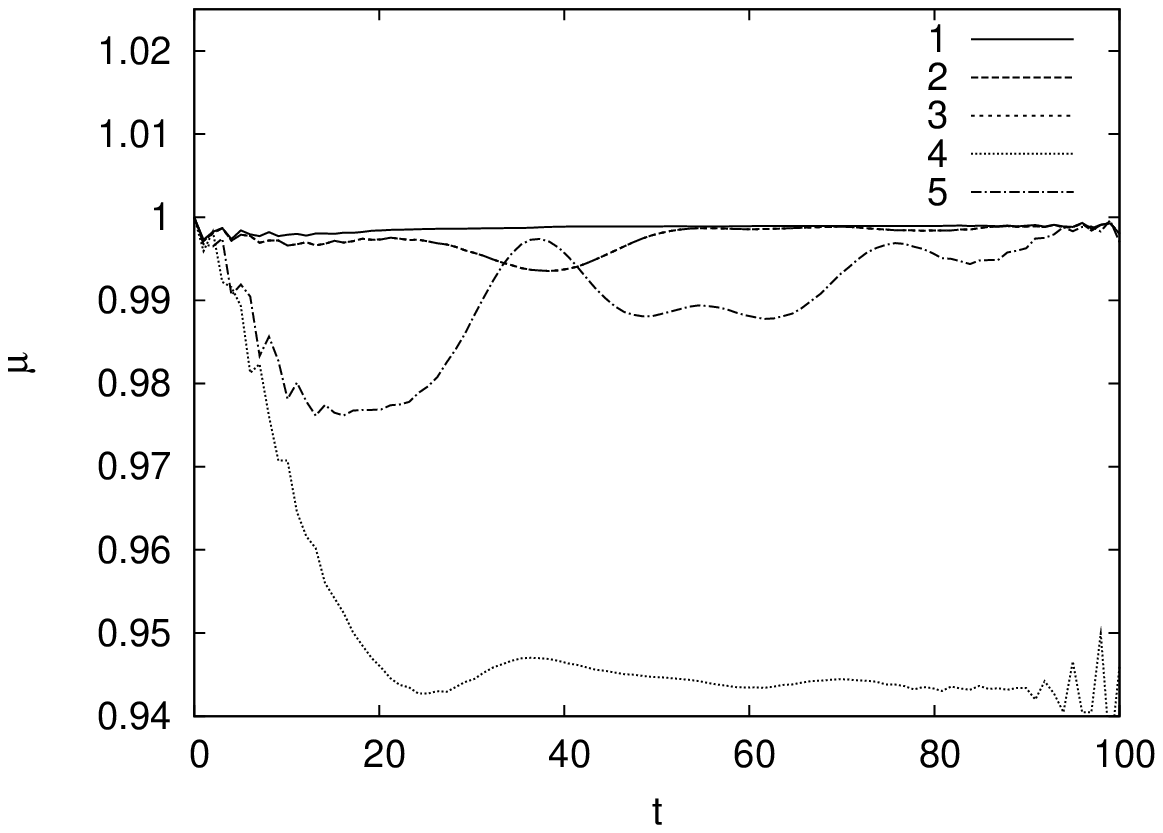}
\includegraphics[width=8. cm]{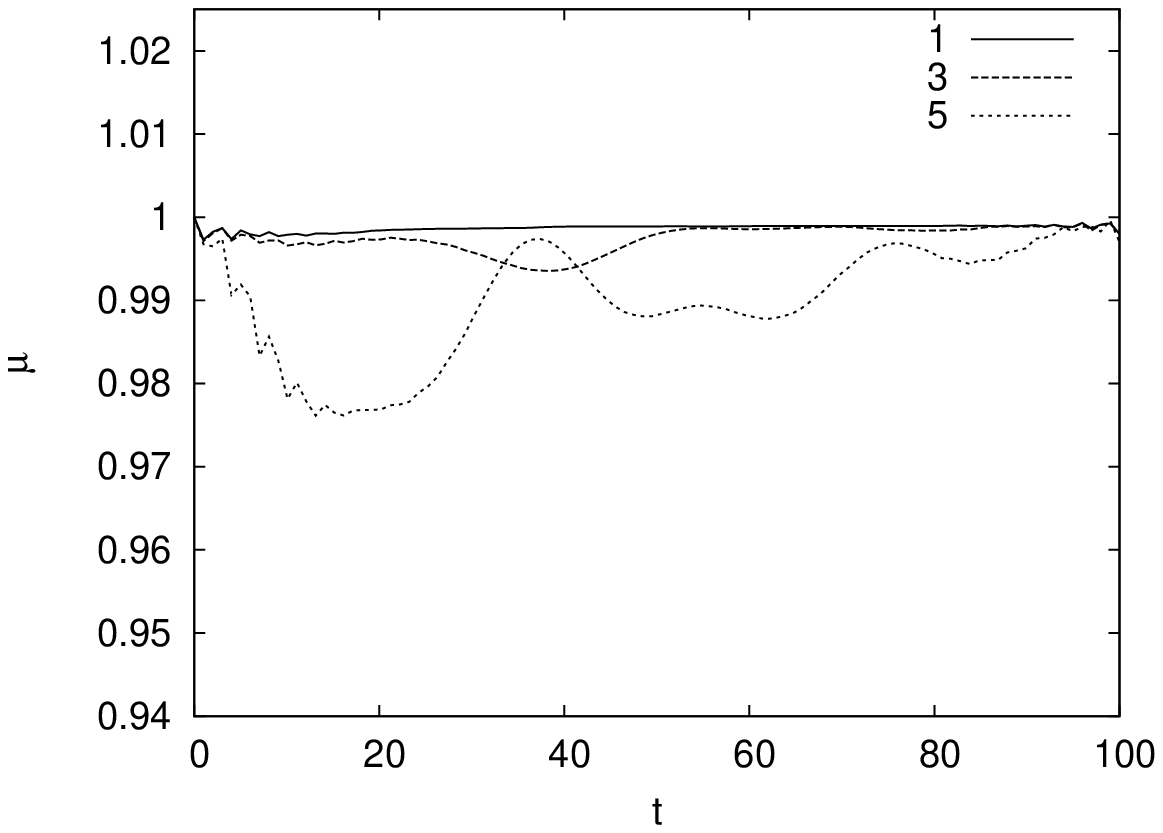}
\caption{The calculated purity $\mu$ at different times $t$ for $\lambda=0.1$, case A,
bath parameters $BP(1,...5)$  (left panels) and $BP(1,3,5)$ (right panels) and number of oscillators $N=11$
(upper panels) and $N=101$ (lower panels). The $i$-th curve corresponds to $BP(i)$}
\label{fig2}
\end{center}
\end{figure}

We first consider case A. Recall that then at $t=0$ $R_{12}=0$ and $\mu(0)=1$, so that the initial state of system 0 is pure.
In Fig. \ref{fig2} we show in the left panel the averaged $\mu(t)$ for the set of $BP(i)$, $i=1,..5$ and $N=11$ and 101. In the right panel
we show the averaged $\mu(t)$ for the restricted set of $BP(i)$, $i=1,3,5$ and the same values of $N$.

\begin{figure}
\begin{center}
\includegraphics[width=8. cm]{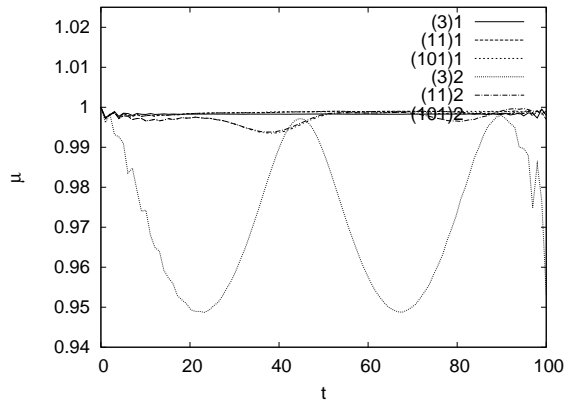}
\caption{The calculated purity $\mu$ at different times $t$  for $\lambda=0.1$, case A,
bath parameters $BP(1,2)$ and number of oscillators $N=3$, 11 and 101. Curve $(N)i$
corresponds to $BP(i)$ with $N$ oscillators.}
\label{fig3}
\end{center}
\end{figure}

For illustration of the notion of the bath
in Fig. \ref{fig3}  we also present the results including $N=3$ when the "bath" is reduced to a
pair of oscillators. In this case only bath parameters $BP(1,2)$ are possible.

To characterize the bath dependence one may use the dispersion of the values of $\mu$ obtained
by using different initial bath data. If we denote $\mu_i$ as the purity obtained from the bath data $BP(i)$
then the width of the dispersion in the interval of times $t\in\Delta t$ can be defined as
\beq
w=\max_{t\in\Delta t}\max_{i\neq j}|\mu_i-\mu_j|,
\label{w}
\eeq
where $i,j$ belong to a set of initial conditions $BP(i)$ used to find the width.
In this way for each $N$  we obtain three different widths $w^{(2)}(N)$, $w^{(3)}(N)$ and $w^{(5)}(N)$ from the sets $BP(1,2)$,
$BP(1,3,5)$ and $BP(1,...5)$.
In the Table 1. we show values of the width for 5 intervals $\Delta t$ in the interval $0<t<100$.
\vspace*{0.2 cm}
\begin{table}
\begin{center}
{\bf Table 1 ($\mu(0)=1$})\\
\vspace*{0.5 cm}
\begin{tabular}{|l|r|r|r|r|r|}
\hline
$\Delta t$&$w^{(2)}$(3)&$w^{(3)}$(11)&$w^{(3)}$(101)&$w^{(5)}$(11)&$w^{(5)}$(101)\\\hline
[0,20]&0.0615&0.0326&0.0381&0.0655&0.0655\\\hline
[20,40]&0.0613&0.0414&0.0387&0.0762&0.0793\\\hline
[40,60]&0.0591&0.0235&0.0206&0.0707&0.0718\\\hline
[60,80]&0.0618&0.0209&0.232&0.0306&0.0720\\\hline
[80,100]&0.0468&0.0171&0.00898&0.0563&0.0709\\\hline
\end{tabular}
\end{center}
\end{table}

\vspace*{0.2 cm}
{\bf Case B}
\vspace*{0.2 cm}

Now we consider case B when at $t=0$ $R_{12}=0.325$ and $\mu=0.592$, so that the initial state
of system 0 is mixed.

\begin{figure}
\begin{center}
\includegraphics[width=8. cm]{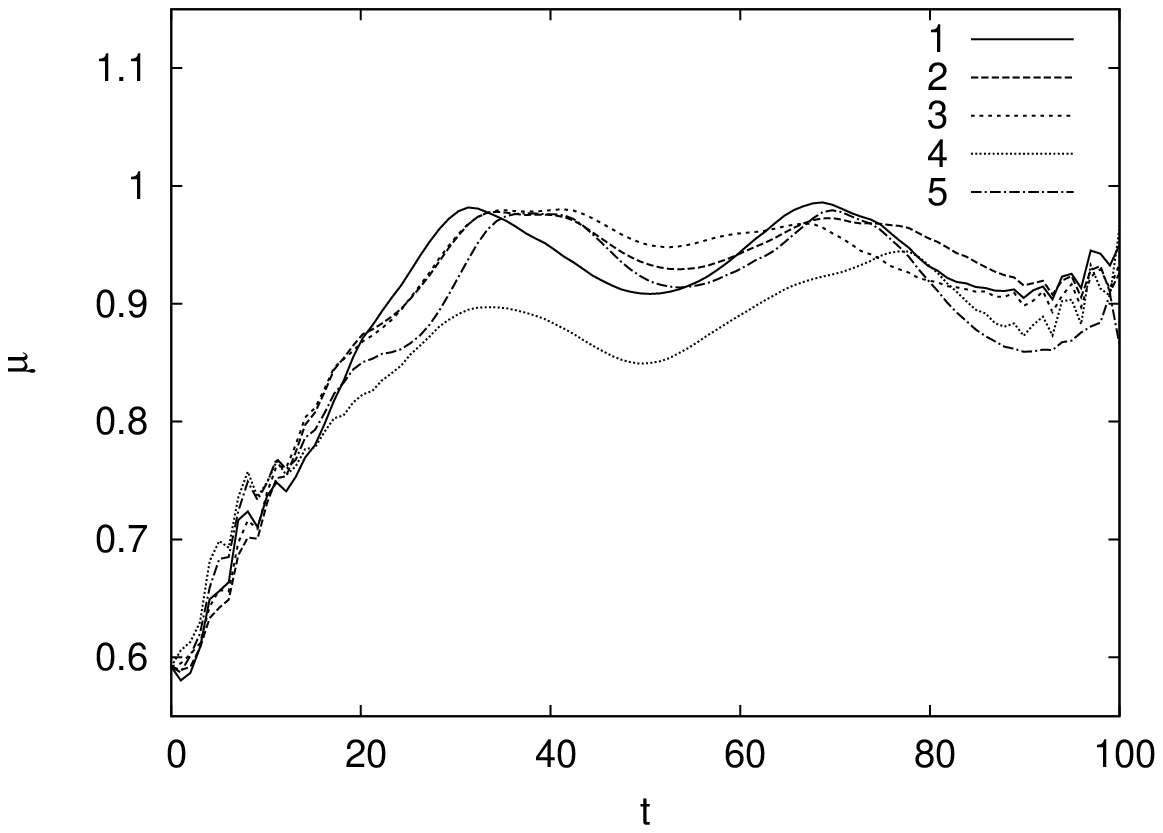}
\includegraphics[width=8. cm]{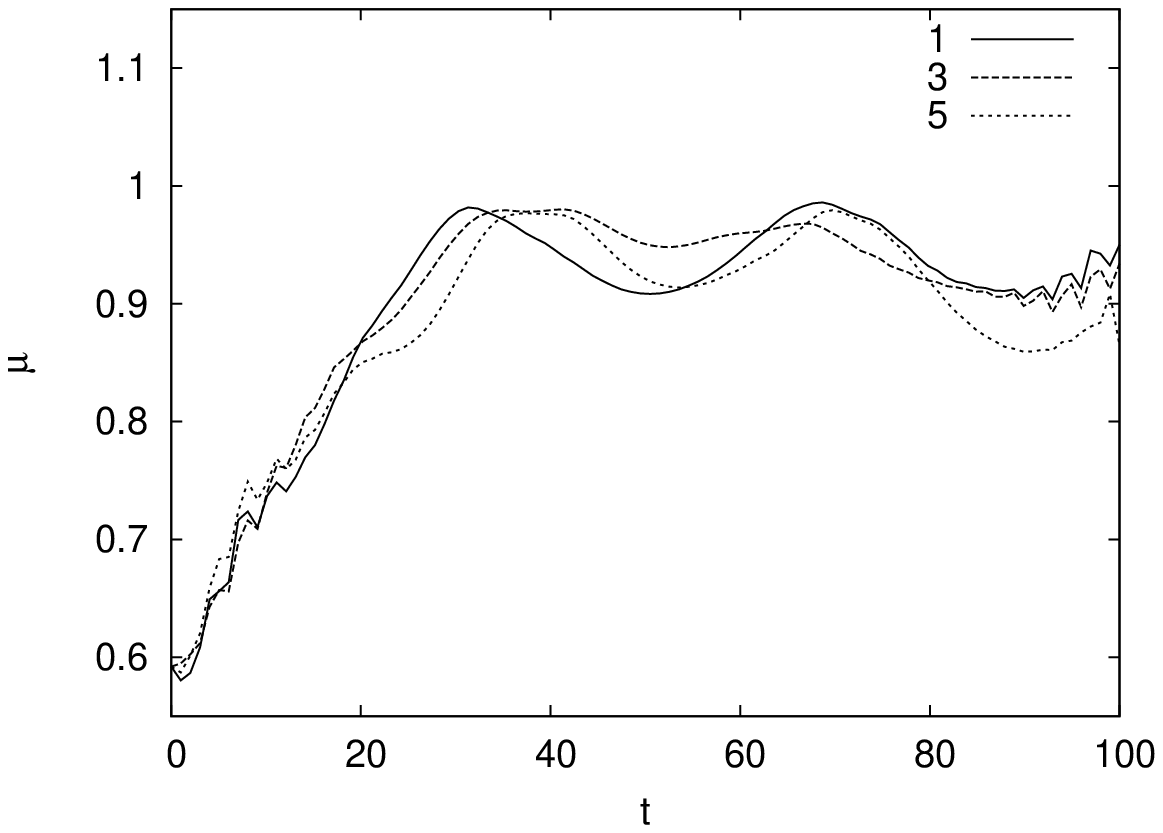}
\includegraphics[width=8. cm]{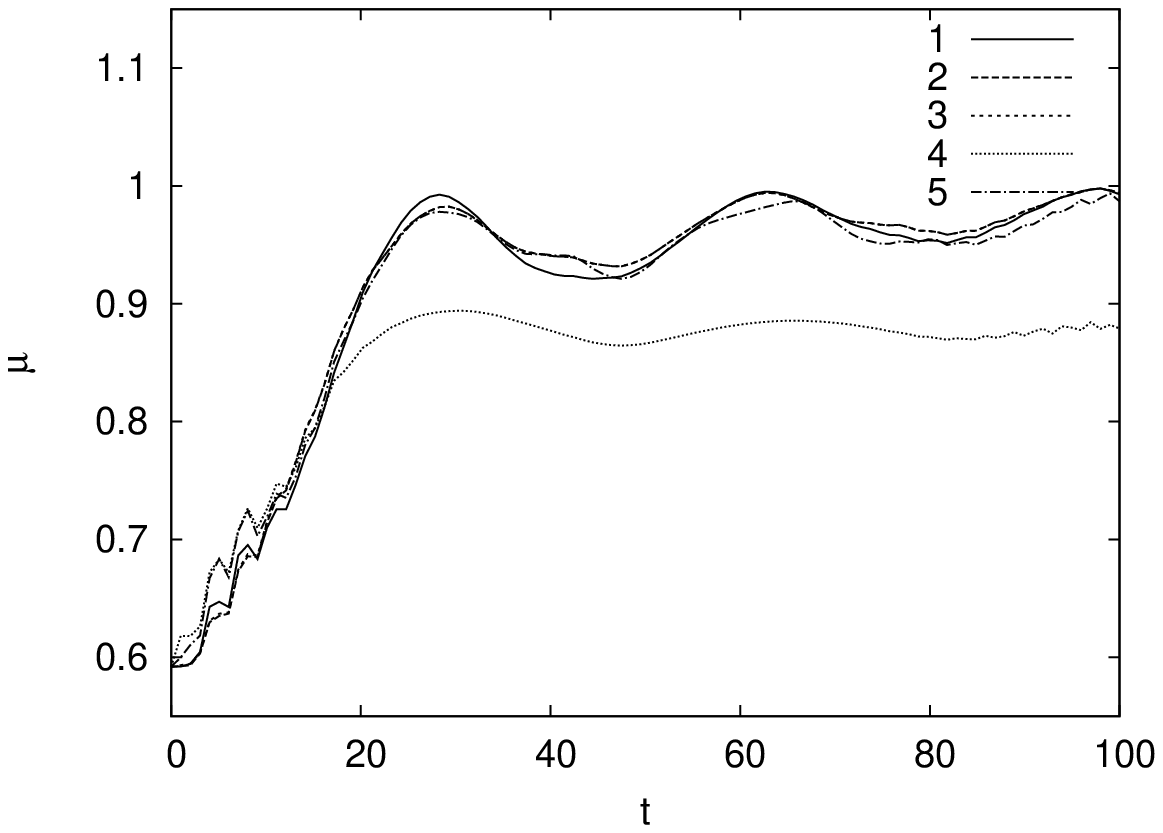}
\includegraphics[width=8. cm]{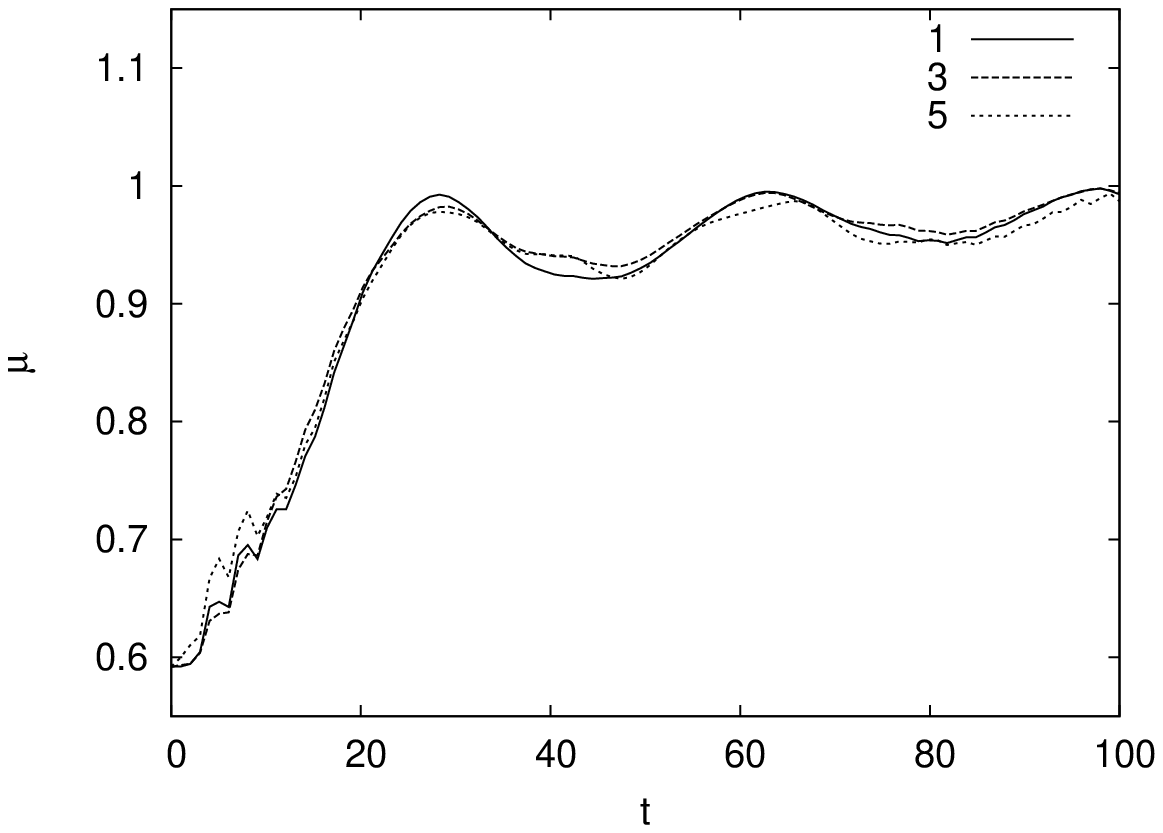}
\caption{The calculated purity $\mu$ at different times $t$ for $\lambda=0.1$, case B,
bath parameters $BP(1,...5)$  (left panels) and $BP(1,3,5)$ (right panels) and number of oscillators $N=11$
(upper panels) and $N=101$ (lower panels). The $i$-th curve corresponds to $BP(i)$}
\label{fig4}
\end{center}
\end{figure}

In Fig. \ref{fig4} as before
we show in the left panel the averaged $\mu(t)$ for the set of $BP(i)$, $i=1,..5$ and $n=11$ and 101. In the right panel
we show the averaged $\mu(t)$ for the restricted set of $BP(i)$, $i=1,3,5$ and the same values of $N$.

\begin{figure}
\begin{center}
\includegraphics[width=8. cm]{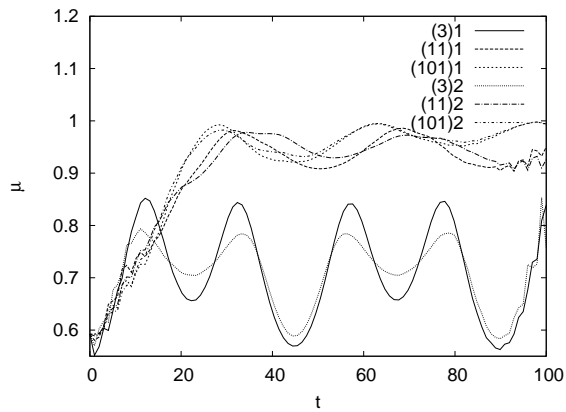}
\caption{The calculated purity $\mu$ at different times $t$ for $\lambda=0.1$, case B,
bath parameters $BP(1,2)$ and number of oscillators $N=3$, 11 and 101.
Curve $(N)i$
corresponds to $BP(i)$ with $N$ oscillators.}
\label{fig5}
\end{center}
\end{figure}

For illustration of the notion of the bath
in Fig. \ref{fig5}  we again present the results including $N=3$ when the "bath" is reduced to a
pair of oscillators. In this case only bath parameters $BP(1,2)$ are possible.

Similarly to case A we present the widths of the dispersion (\ref{w})
in Table 2.
\vspace*{0.2 cm}
\begin{center}
{\bf Table 2 ($\mu(0)=0.592$})\\
\vspace*{0.5 cm}
\begin{tabular}{|l|r|r|r|r|r|}
\hline
$\Delta t$&$w^{(2)}$(3)&$w^{(3)}$(11)&$w^{(3)}$(101)&$w^{(5)}$(11)&$w^{(5)}$(101)\\\hline
[0,20]&0.247&0.132&0.131&0.182&0.155\\\hline
[20,40]&0.261&0.105&0.0577&0.125&0.121\\\hline
[40,60]&0.170&0.0930&0.0412&0.180&0.133\\\hline
[60,80]&0.275&0.0782&0.0364&0.157&0.124\\\hline
[80,100]&0.177&0.112&0.0278&0.128&0.128\\\hline
\end{tabular}
\end{center}

\subsection{Numerical results, $\lambda=1.0$}

\begin{figure}
\begin{center}
\includegraphics[width=8. cm]{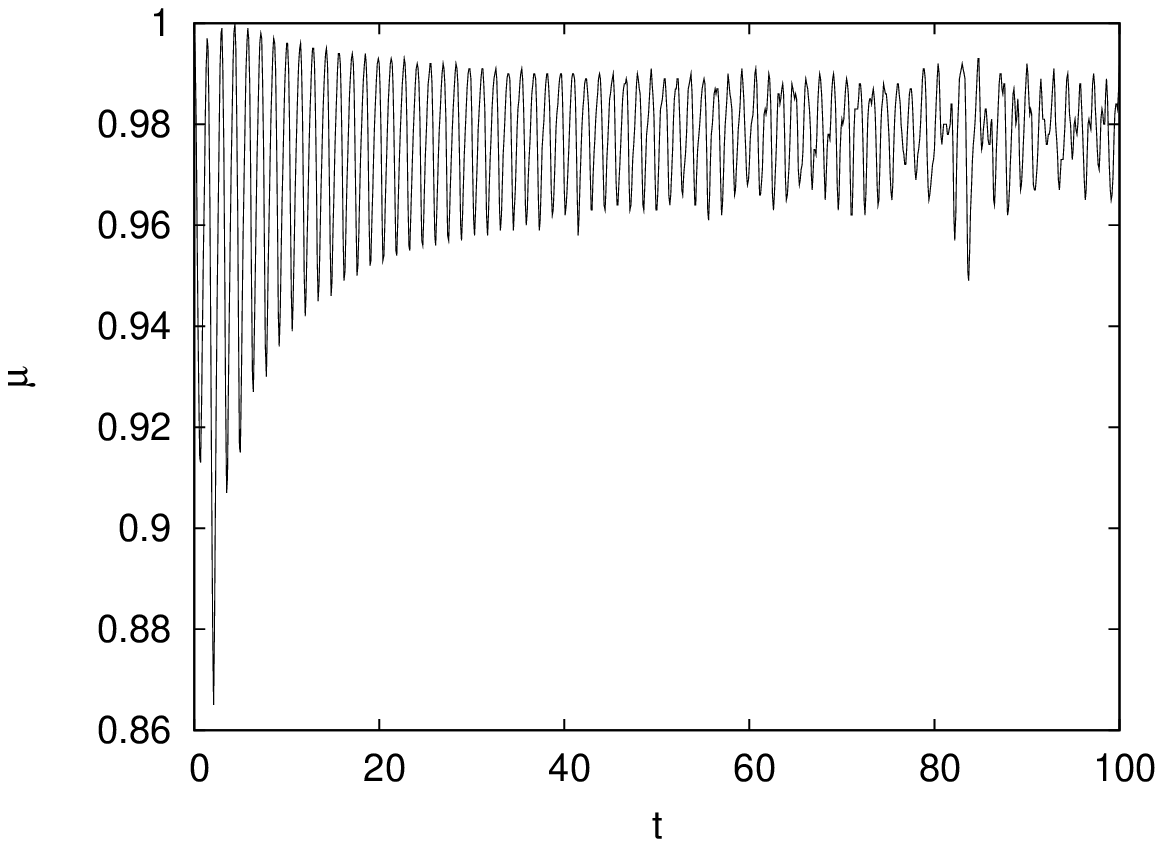}
\includegraphics[width=8. cm]{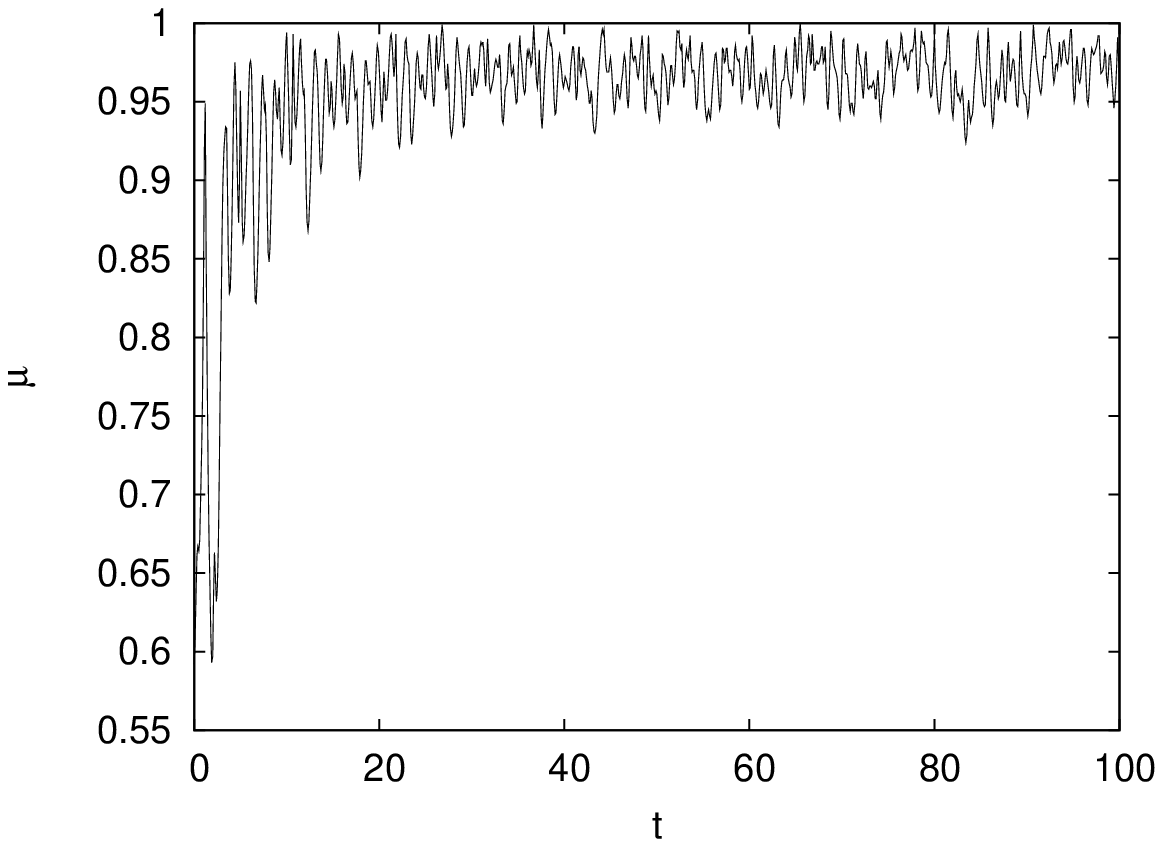}
\caption{The calculated purity $\mu$ at different times $t$ for $\lambda=1.0$, cases A(left panel) and B( right panel),
bath parameter $BP(1)$ and number of oscillators $N=101$}
\label{fig6}
\end{center}
\end{figure}

With $\lambda=1.0$ the average frequency grows roughly by 50\%  and so the number of oscillations in the interval from $t=0$ up to $t=100$.
Comparing to the previous case with $\lambda=0.01$ this time interval roughly corresponds to a shorter one $0<t<67$.
This is illustrated in Fig. \ref{fig6} in which we show natural values of $\mu(t)$ for $BP(1)$ and
both cases A and B with $N_1=100$.
As before in the following we present the purity  averaged over short range oscillation by
the Bezier procedure.\\

\vspace*{0.2 cm}
{\bf Case A}\\
\vspace*{0.2 cm}

\begin{figure}
\begin{center}
\includegraphics[width=8. cm]{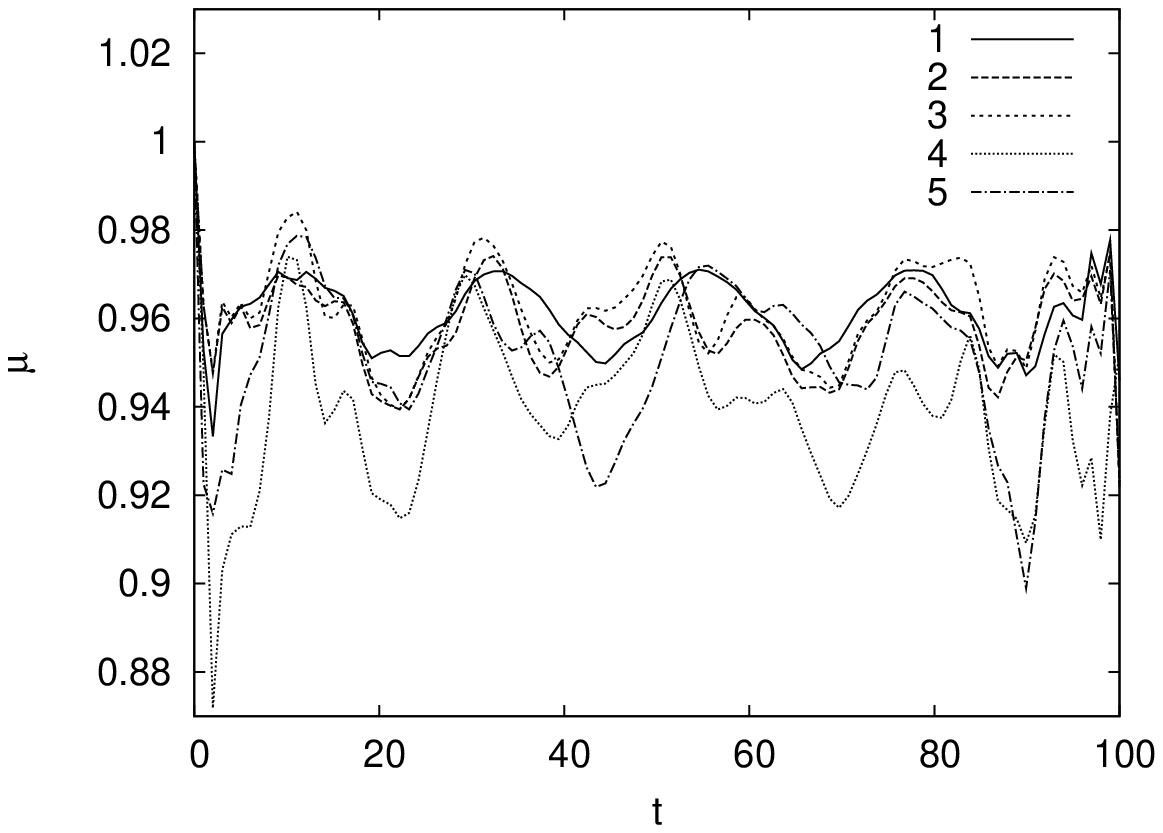}
\includegraphics[width=8. cm]{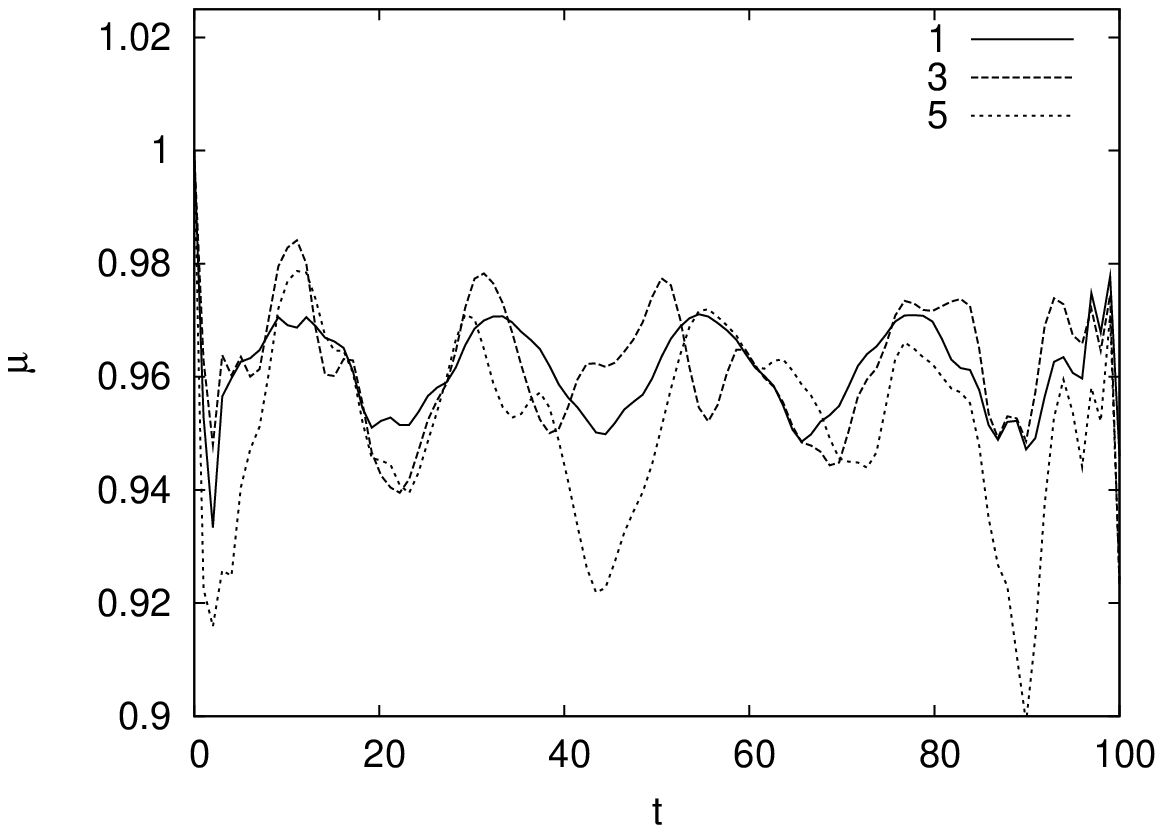}
\includegraphics[width=8. cm]{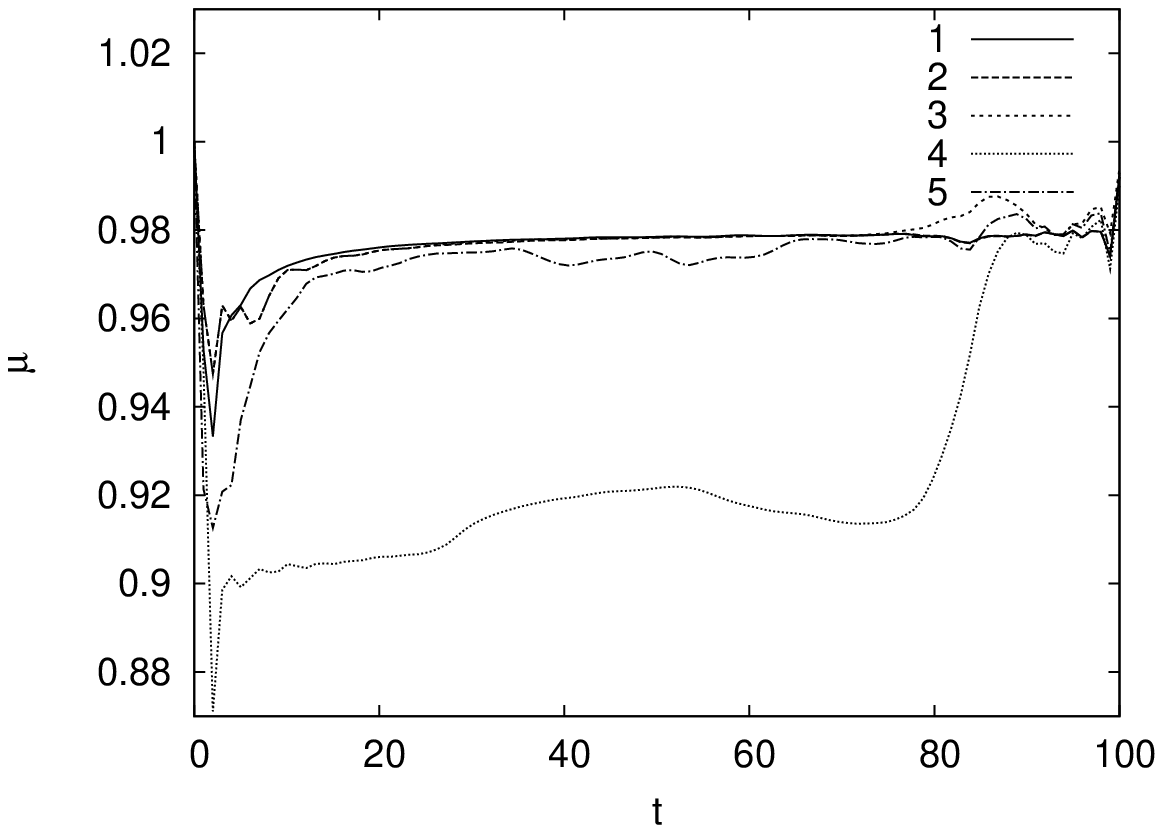}
\includegraphics[width=8. cm]{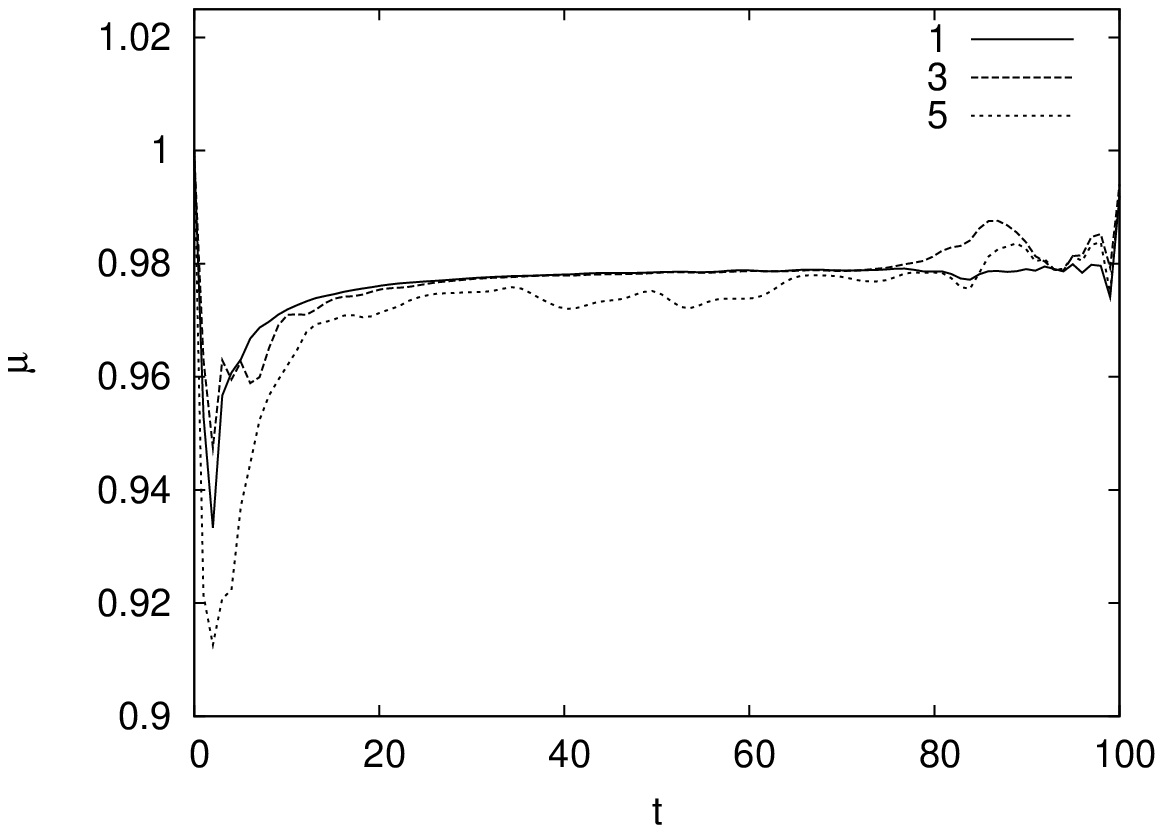}
\caption{The calculated purity $\mu$ at different times $t$ for $\lambda=1.0$, case A,
bath parameters $BP(1,...5)$  (left panels) and $BP(1,3,5)$ (right panels) and number of oscillators $N=11$
(upper panels) and $N=101$ (lower panels). The $i$-th curve corresponds to $BP(i)$}
\label{fig7}
\end{center}
\end{figure}

Again we start with case A. Recall that then at $t=0$ $R_{12}=0$ and $\mu(0)=1$, so that the initial state of system 0 is pure.
In Fig. \ref{fig7} we show in the left panel the averaged $\mu(t)$ for the set of $BP(i)$, $i=1,..5$ and $N=11$ and 101. In the right panel
we show the averaged $\mu(t)$ for the restricted set of $BP(i)$, $i=1,3,5$ and the same values of $N$.

\begin{figure}
\begin{center}
\includegraphics[width=8. cm]{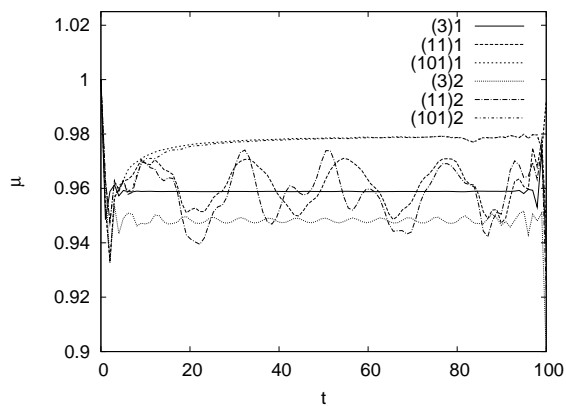}
\caption{The calculated purity $\mu$ at different times $t$ for $\lambda=1.0$, case A(left panel),
bath parameters $BP(1,2)$ and number of oscillators $N=3$, 11 and 101.
 Curve $(N)i$
corresponds to $BP(i)$ with $N$ oscillators.}
\label{fig8}
\end{center}
\end{figure}

For illustration of the notion of the bath
in Fig. \ref{fig8}  we once more present the results including $N=3$ when the "bath" is reduced to a
pair of oscillators. In this case only bath parameters $BP(1,2)$ are possible.

The width $w$ defined by (\ref{w})
is shown in Table 3. Recall that  three different widths $w^{(2)}(N)$, $w^{(3)}(N)$ and $w^{(5)}(N)$ correspond to the sets $BP(1,2)$,
$BP(1,3,5)$ and $BP(1,...5)$. As before we  show values of the width for 5 intervals $\Delta t$ in the interval $0<t<100$.
\vspace*{0.2 cm}
\begin{table}
\begin{center}
{\bf Table 3 ($\mu(0)=1$})\\
\vspace*{0.5 cm}
\begin{tabular}{|l|r|r|r|r|r|}
\hline
$\Delta t$&$w^{(2)}$(3)&$w^{(3)}$(11)&$w^{(3)}$(101)&$w^{(5)}$(11)&$w^{(5)}$(101)\\\hline
[0,20]&0.0848&0.120&0.117&0.161&0.154\\\hline
[20,40]&0.0848&0.104&0.0239&0.141&0.120\\\hline
[40,60]&0.0846&0.215&0.0241&0.215&0.0946\\\hline
[60,80]&0.0848&0.0967&0.0123&0.115&0.0992\\\hline
[80,100]&0.0845&0.206&0.0226&0.224&0.0801\\\hline
\end{tabular}
\end{center}
\end{table}

\vspace*{0.2 cm}
{\bf Case B}
\vspace*{0.2 cm}

Now we consider case B when at $t=0$ $R_{12}=0.325$ and $\mu=0.592$, so that the initial state
of system 0 is mixed.

\begin{figure}
\begin{center}
\includegraphics[width=8. cm]{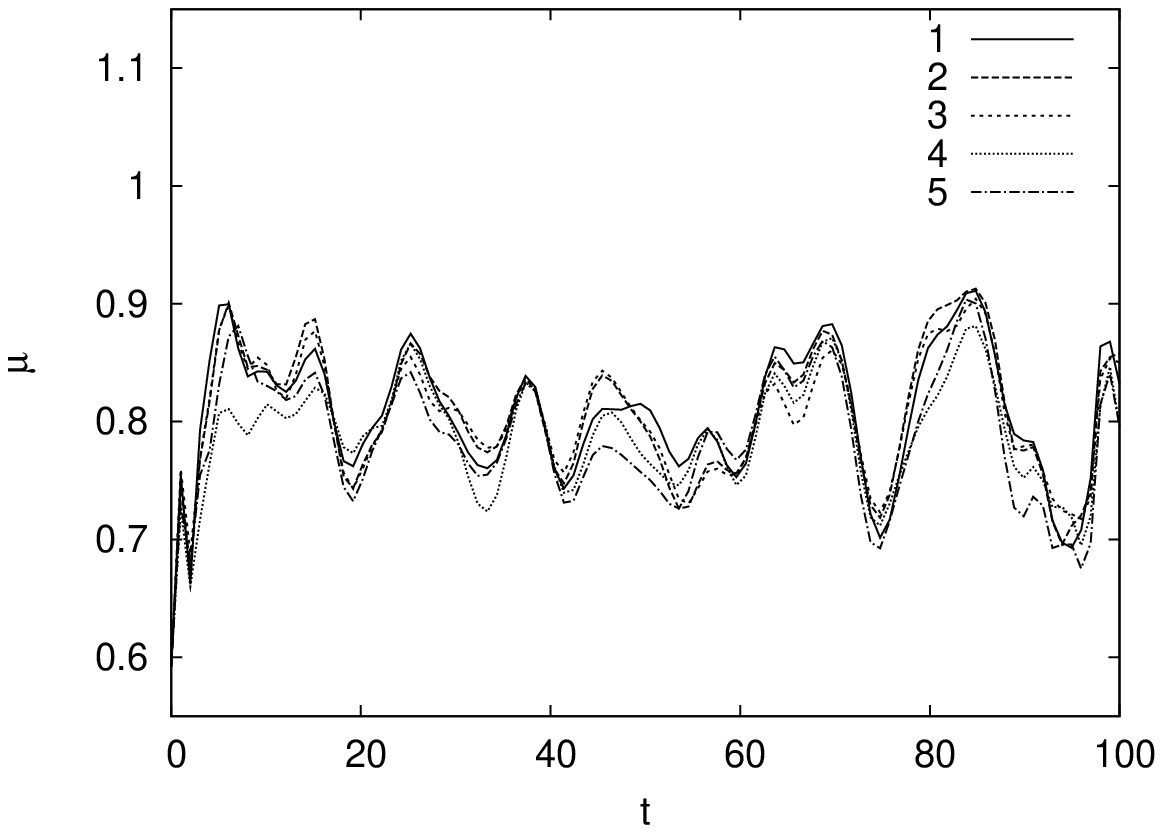}
\includegraphics[width=8. cm]{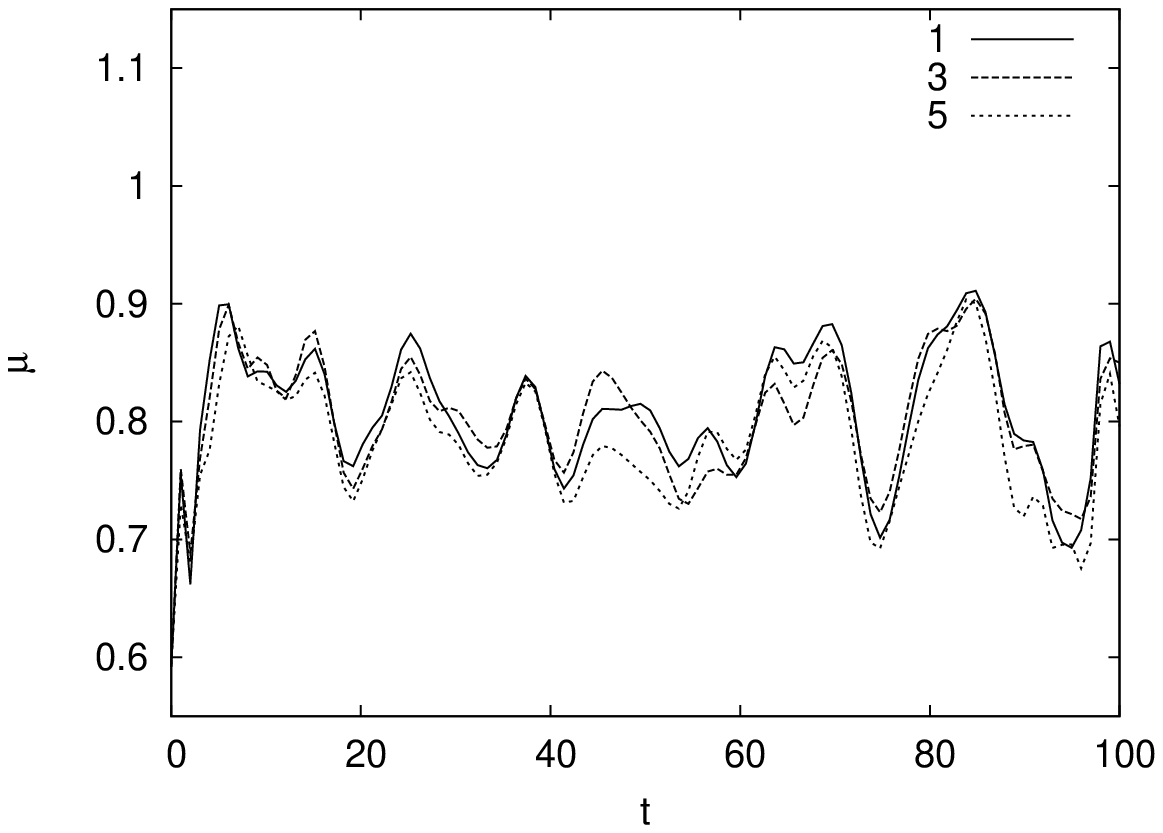}
\includegraphics[width=8. cm]{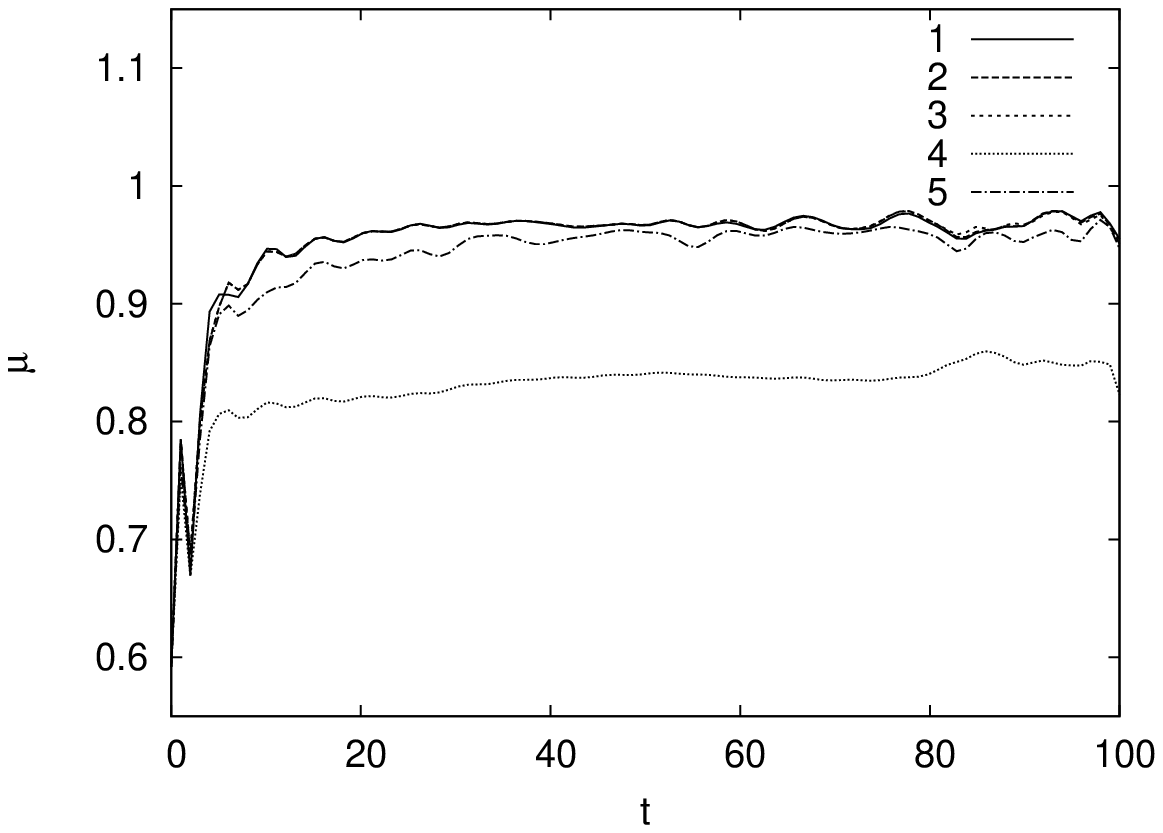}
\includegraphics[width=8. cm]{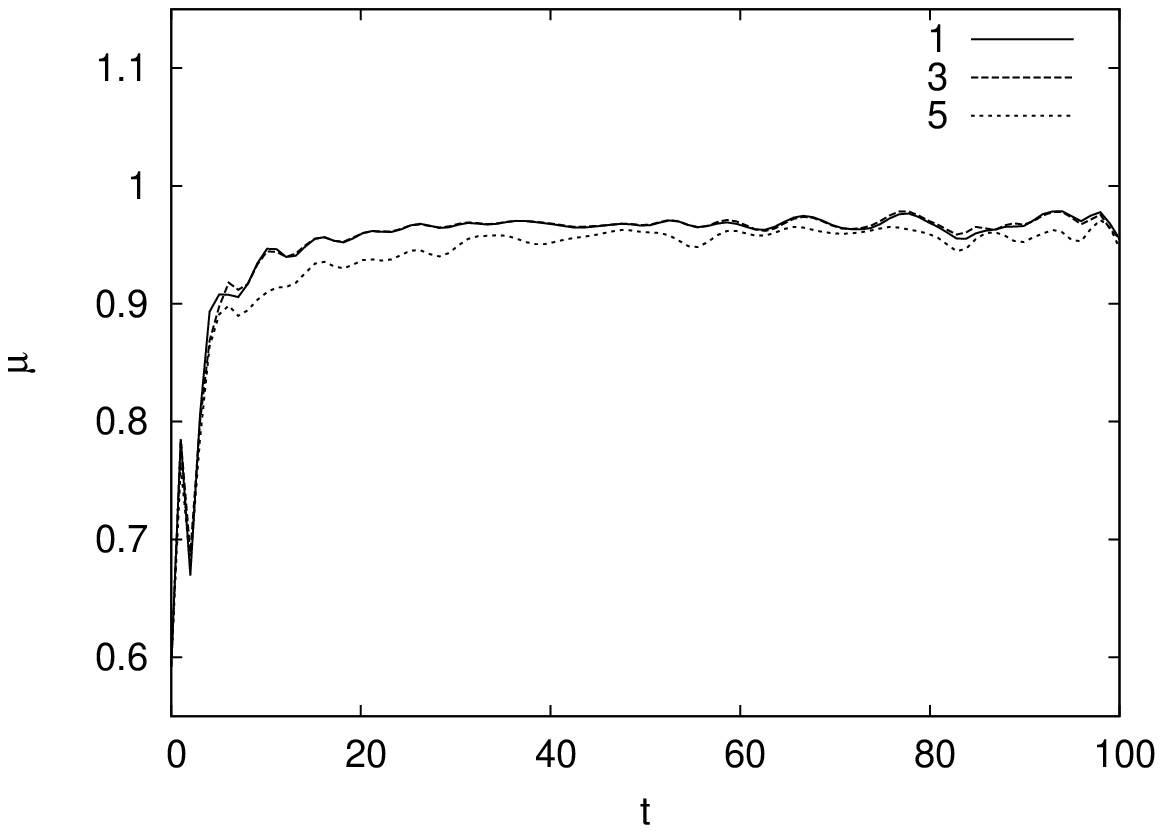}
\caption{The calculated purity $\mu$ at different times $t$ for $\lambda=1.0$, case B,
bath parameters $BP(1,...5)$  (left panels) and $BP(1,3,5)$ (right panels) and number of oscillators $N=11$
(upper panels) and $N=101$ (lower panels). The $i$-th curve corresponds to $BP(i)$}
\label{fig9}
\end{center}
\end{figure}

In Fig. \ref{fig9}
we show in the left panel the averaged $\mu(t)$ for the set of $BP(i)$, $i=1,..5$ and $n=11$ and 101. In the right panel
we show the averaged $\mu(t)$ for the restricted set of $BP(i)$, $i=1,3,5$ and the same values of $N$.

\begin{figure}
\begin{center}
\includegraphics[width=8. cm]{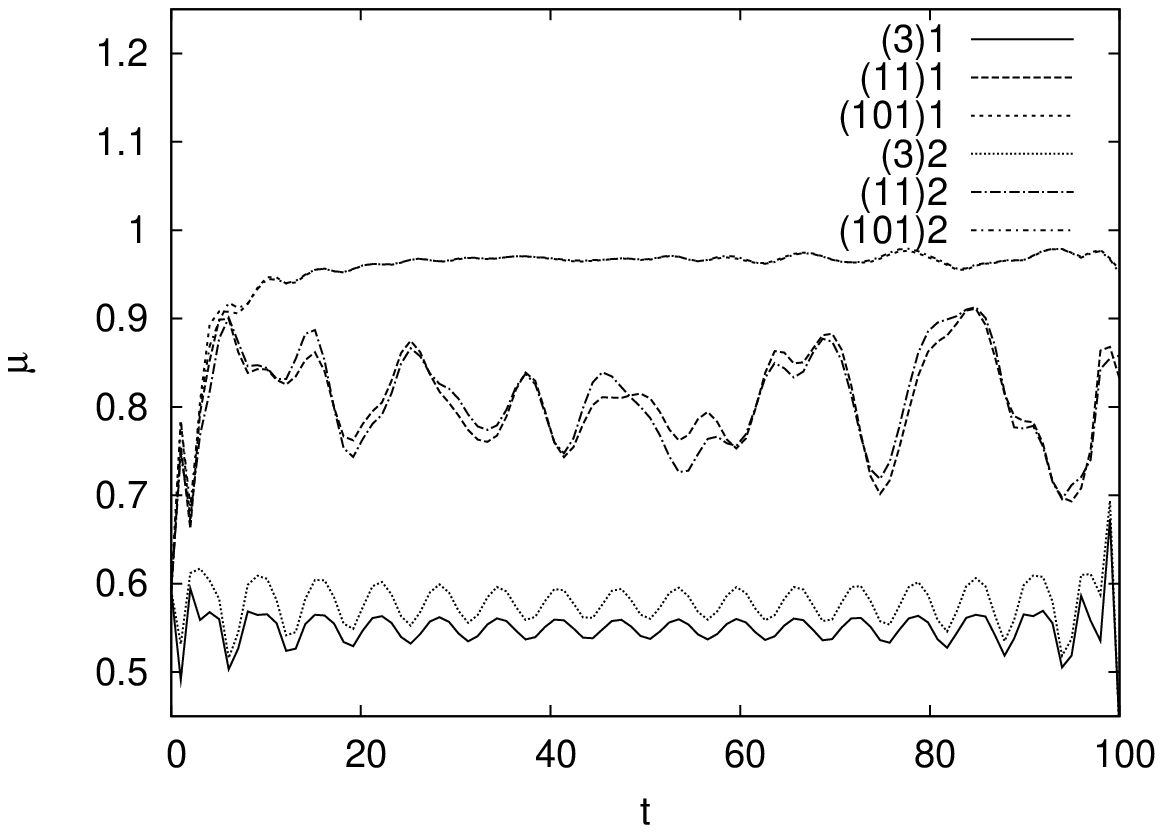}
\caption{The calculated purity $\mu$ at different times $t$ for $\lambda=1.0$, case B,
bath parameters $BP(1,2)$ and number of oscillators $N=3$, 11 and 101.
 Curve $(N)i$
corresponds to $BP(i)$ with $N$ oscillators.}
\label{fig10}
\end{center}
\end{figure}

To illustrate the notion of the bath with its strong coupling $\lambda=1.0$
in Fig. \ref{fig10}  we again present the results including $N=3$ when the "bath" is reduced to a
pair of oscillators. In this case only bath parameters $BP(1,2)$ are possible.

The widths of the dispersion (\ref{w}) for case B and $\lambda=1.0$ are presented
in Table 4\\
\vspace*{4 cm}
\begin{center}
{\bf Table 4 ($\mu(0)=0.592$})\\
\vspace*{0.5 cm}
\begin{tabular}{|l|r|r|r|r|r|}
\hline
$\Delta t$&$w^{(2)}$(3)&$w^{(3)}$(11)&$w^{(3)}$(101)&$w^{(5)}$(11)&$w^{(5)}$(101)\\\hline
[0,20]&0.201&0.198&0.136&0.234&0.183\\\hline
[20,40]&0.201&0.146&0.0602&0.198&0.170\\\hline
[40,60]&0.201&0.161&0.0543&0.200&0.156\\\hline
[60,80]&0.201&0.170&0.0450&0.187&0.160\\\hline
[80,100]&0.200&0.206&0.0440&0.235&0.151\\\hline
\end{tabular}
\end{center}

\section{Discussion}
We start with the weak coupling to the bath usually assumed for the evolution ("the Born-Markov approximation" ~\cite{brasil,lidar}).
Inspection of our results demonstrated in Figs. \ref{fig1} -\ref{fig5} and Tables 1. and 2. leads to the following conclusions.

In all cases evolution of the purity for the $\rho$-matrix of the subsystem (system 0) is not Markovian.
However deviations from the Markov evolution substantially depend on the property of the bath and elapsed time.

If the bath is composed of only two other oscillators 1 and 2  then deviations from the Markov evolutions are quite strong
and practically persist at all times from $t=0$ to $t=100$. The scale of deviations measured by the width $w$ is of the order 50\%.

With the growth of the number of oscillators composing the bath the deviation from the Markov behavior diminish and they also diminish with time.
This is especially visible if the bath change does not directly involves the interaction with the subsystem, that is for bath parameters $BP(1)$, $BP(3)$
and $BP(5)$. In this case  for the bath with 10 oscillators the deviations from the Markov evolution fall to 10\% . With the bath composed of 100 oscillators
the deviations fall from 10\% at $0<t<20$ to 3\% at $80<t<100$. So for such a "macroscopic" bath one can assume  validity of the Markov
 behavior at large enough times.

Inclusion of bath changes at the direct interaction with a subsystem spoils this nice picture and returns to the deviations of the order
15\%-30\%, which only very slowly diminish with time. So in this case the evolution is definitely non-Markovian.

One of the unexpected conclusions of our numerical exercise is that for the bath composed of 10 or 100 oscillators the initially mixed state
tends to the pure one with a growth of time. This growth is quite fast: already at $t=30$ the state becomes practically pure and remains such at
later times.

With a raised coupling constant, that is with a  strong bath-system interaction,  from Figs.\ref{fig6}-\ref{fig10} and Tables 3 and 4
we may conclude that on the general the evolution is of the same pattern as with a weak interaction, with  raised widths.
So the deviations from the markovian evolution become stronger. Also these deviations do not substantially diminish with time.
The purity of the system 0 in this  case does not generally become restored at large times. With the initial purity $\mu(0)=0.592$  it freezes at the order 0.8.
A remarkable exception is the behavior with the bath of 101 oscillators  without direct contact with the system 0.  This behavior  is not very different
from the weak coupling case and leads to the nearly markovian evolution at long times with  somewhat raised widths.   Also in this case the system
tends to the pure one, as with a weak coupling.

So in conclusion one finds that the assumption of the Markovian behavior
is not a bad approximation provided the bath
has a "macroscopic" character, is randomly connected with the subsystem and the time of evolution is long enough.
This is true both for weak and strong coupling of the system with the bath.

Note that in ~\cite{lidar,nathan} certain criteria for the Markov evolution of the $\rho$ matrix of the open system were
presented in the form of estimations of the error of the corresponding Lindblad equation. Unfortunately they refer to the case when the bath
is described by a bounded operator and so cannot be applied to our bath of a set of oscillators.
As an alternative in ~\cite{lidar} a new coarse grained Lindblad-like equation was proposed for the $\rho$ matrix averaged over some
prescribed intervals of time together  with the corresponding error bounds. This time averaging probably is equivalent to our Bezier procedure
used for plotting our results but it has no relation to our widths read from the results without any averaging.  So again it is very difficult
to compare the degree of non-Markovianity obtained from this pure theoretical derivation and our numerical results. Note that the errors found in
~\cite{lidar} in all cases grow exponentially with the time of evolution. In contrary our widths in the most beneficial case (with the bath better corresponding
to its standardly assumed properties)  steadily diminish with time indicating restoration of markovianity at large enough times as stressed above.
So probably the error bounds presented in ~\cite{lidar} were too stringent and the applicability of the Lindblad equation has a wider time limitation.

We have to stress that our conclusions have been derived only from the behavior of the purity, which of course is only one of the properties of the
$\rho$-matrix for the subsystem. Still we think that the purity gives a very conclusive manifestation of the global behavior of this $\rho$ matrix.
We cannot exclude that there exist some observables which are more sensitive to the bath parameters and so not described by the Markovian evolution.
It is not easy to pinpoint such observables apriori. They can be found only in the study of some concrete problems, which we postpone for
future studies.

\end{document}